\documentclass{aa}
\usepackage{longtable,lscape}
\usepackage{graphicx}
\usepackage{lscape}

\usepackage{times}
\usepackage{epsfig}
\usepackage{epsf}
\usepackage{rotating}

\begin{document}


\title{New 3-D gas density maps of NaI and CaII interstellar absorption within 300pc. \footnote{Partly based on observations collected at the European Southern Observatory, La Silla, Chile.}}


\author{
Barry Y. Welsh\inst{1},
 Rosine Lallement \inst{2}, Jean-Luc Vergely \inst{3} and S\'everine Raimond \inst{2}}

\institute{Space Sciences Laboratory, University of California, 7 Gauss Way, Berkeley, CA 94720 \and 
Universit\'e Versailles St-Quentin; CNRS/INSU, LATMOS-IPSL , B.P.3, 91371 Verri\`eres-le-Buisson, France \and ACRI-ST, BP 234, 06504 Sofia-Antipolis, France}

\date{Started: April 2008}

\titlerunning{NaI and CaII maps}
\authorrunning{Welsh $\&$ Lallement}


\abstract
{}
{We present new high resolution (R $>$ 50,000) absorption measurements of the NaI doublet (5889 - 5895\AA) 
along 482 nearby sight-lines, in addition to 807 new measurements of the CaII K (3933\AA)
absorption line. We have combined these new data with previously reported measurements to
produce a catalog of absorptions towards a total of 1857 early-type stars located within 800pc of the Sun. 
Using these data we have determined the approximate 3-dimensional spatial distribution of neutral and partly ionized interstellar gas
density within
a distance-cube of 300pc from the Sun. }
{All newly recorded spectra were analyzed by means of a multi-component line profile-fitting program, in most
cases using simultaneous fits to the line doublets. Normalized absorption profiles were fitted by varying the velocity,
doppler width and column density for all intervening interstellar clouds. The resulting total column densities were
then used in conjunction with the Hipparcos distances of the target stars to construct inversion maps of the
3-D spatial density distribution of the NaI and CaII bearing gas.}
{A plot of the equivalent width of NaI versus distance reveals a wall of neutral gas at $\sim$80pc that can be associated
with the boundary wall to the central rarefied Local Cavity region. In contrast, a similar plot for the equivalent width of CaII shows no sharply increasing absorption at 80pc, but instead we observe a slowly increasing value of CaII  equivalent width with increasing sight-line distance sampled.
}
{}



\maketitle


\section{Introduction}
Knowledge of the spatial
distribution, dynamics and the associated physical and chemical state of  interstellar gas can
provide important insights into
the subtle interplay between the evolution of stars and their exchange of material to and from
the ambient interstellar medium (ISM). Although much valuable work in this field
has been achieved through numerous 21cm radio and infrared surveys of
our Galaxy, knowledge of
the physical state and spatial distribution of
of the interstellar gas within 300pc of the Sun is still far from complete. The low
density region surrounding our Sun to $\sim$ 100pc in all directions is termed
`the Local Cavity' (LC) and has been shown to be largely free of cold and dense
gas, although it does contain many partially ionized diffuse cloudlets often referred to
as `local fluff' (Redfield $\&$ Linsky \cite{redfield08}). The actual physical state
of the gas that presumably fills the remaining
volume in the Local Cavity is still much debated. The prevailing view
is that it is mainly filled with a very low density, highly ionized (million K) gas, as inferred
by observations of the diffuse soft X-ray (0.25keV) background emission (Snowden et al. \cite{snow98})
and also by the detection of local absorption from the
high ionization OVI (1032\AA)  line by Savage $\&$ Lehner (\cite{savage06}). 
However, several authors have
raised some serious arguments against the prevailing interpretation of these data, such
that the LC gas may be of a far lower temperature or in a state far from
ionization equilibrium (Breitschwerdt \cite{breit01}; Koutroumpa et al. \cite{kout07}; Barstow et al. \cite{barstow08}). 
Based on the demonstration that most of the unabsorbed soft X-ray brightness at low galactic
latitudes is actually solar wind charge-transfer X-ray emission generated within the
heliosphere, Koutroumpa et al.( \cite{kout09}) have suggested that the LC may
contain far cooler or more tenuous gas at low altitudes ($\mid$ $\it z$ $\mid$ $<$ 60 - 100pc) with
the soft X-ray emission seen from $\mid$ $\it b$ $\mid$ $>$ 30$^{\circ}$ arising
from an `external' hot gas and from the overlying halo (Welsh $\&$ Shelton \cite{welsh09}).

Our unique placement within this low density interstellar
cavity allows us to directly measure its spatial extent and
physical state through high spectral resolution absorption measurements that are
mostly uncontaminated by 
intervening dense line-of-sight interstellar features.
Over the past 20+ years we
have been carrying out high
spectral resolution (R $\sim$ 100,000) absorption observations of the
local neutral and partially ionized interstellar gas
using both the NaI D-line doublet at 5890\AA\ and the CaII-K line at 3933\AA\  (Lallement
et al. \cite{lall86}; Vallerga et al. \cite{vallerga93}; Welsh et al. \cite{welsh94}; Sfeir et al.
\cite{sfeir99};  Lallement et al. \cite{lall03}). These, and many other, observations
have allowed us to construct a picture of the distribution
of neutral gas in the local ISM in which the LC is connected
though interstellar tunnels to other surrounding cavities, as predicted in the model of the ISM
by Cox $\&$ Smith (\cite{cox74}). In particular, maps of the distribution of NaI absorption
have revealed 3 large interstellar features that dominate the appearance of the local ISM
within 200pc. These are: (i) the 50pc diameter and 200pc long extension to the LC 
in the direction of the star $\beta$ CMa (Welsh \cite{welsh91}),
(ii) the extension of the rarefied LC into the lower galactic halo to form an open-ended
Local Chimney feature (Welsh et al. \cite{welsh99}), and (iii) the apparent connection
of the LC with the large Loop I superbubble at a distance of $\sim$ 80pc (Welsh $\&$ Lallement 
\cite{welsh05}). We attribute the name "Loop 1 superbubble" to the cavity centered at longitude 345 deg, although the link between this cavity and the radio feature is far from well understood. It seems likely that the existence of these 3 features is somehow
linked to the formation history of the LC itself.

The need for accurate maps of the 3-D density distribution of neutral local gas has been
demonstrated in wide variety of recent studies (Oegerle et al. \cite{oeg05}; Smith et al. \cite{smith05}; Combet et al. \cite{comb05}; Posselt et al. \cite{poss07}; Wolleben \cite{woll07}). 
Such maps are particularly
relevant for studying the interaction regions between local gas residing in the galactic plane,
gas in adjacent interstellar cavities and local gas being ejected into, and falling from, the lower
galactic halo. It is important to note that $\it only$ when absorption data are combined with
 parallax measurements can the distances to gas clouds be accurately derived. Maps of the
distribution of interstellar gas derived from HI, CO and the IR are of two-dimensions and cannot
provide accurate distance estimates to gas clouds.
Even so,  the accuracy and spatial extent of 
our existing absorption maps are presently limited by (a) the number of sight-line directions sampled,
(b) the sampling length along and between each
sight-line and (c) the distance over which each sight-line extends. At present the
most accurate maps of the density distribution of local neutral gas are those presented by
Lallement et al. (\cite{lall03}). Some 1005 sight-lines within $\sim$ 250pc were
observed using NaI absorption
in order to produce 3-D neutral gas density distribution maps to a distance of $\sim$ 150pc in most directions.
The galactic distribution of these 1005 sight-lines is shown in Figure 1
of Lallement et al. (\cite{lall03}), hereafter Paper 1. Many of these observations
traced the extent of Gould's Belt, since only early-type stars were used as background
absorption sources. We note that in Paper 1
the typical global sampling area on the sky was $\sim$ 1 target per 50 sq. deg.,
with significantly less coverage at high galactic latitudes.

No equivalent maps of the density distribution of partially ionized gas, as
traced by the interstellar CaII absorption line, currently
exist for the local ISM. This is particularly important for sight-lines $\la$ 80pc, since
the LC gas is thought to mainly exist in an ionization state higher than that which
can be probed
by the NaI ion (Welsh et al. \cite{welsh94}). Many diffuse and partly ionized cloudlets
are known to exist within this volume and previous studies have demonstrated
that the CaII K-line is an excellent tracer of absorption of this physical phase of the
local gas (Lallement, Vidal-Madjar $\&$ Ferlet \cite{lall86}; Crawford, Craig $\&$ Welsh \cite{craw97}).
In fact, 
in a comparison study of high resolution ground-based CaII K-line and UV (FeII  and MgII)
absorption line data towards stars within 100pc by
Redfield $\&$ Linsky (\cite{redfield02}), it was found that most (but not all) sight-line components
were observed in both the CaII and UV lines to within a velocity error of only $\pm$1.5 km s$^{-1}$.
This evidence suggests that the CaII line is a good indicator of the absorption structure that one might expect from the warm and
ionized gas sampled by the UV lines. The good agreement between these absorption velocities clearly suggests a co-location for the formation of both the optical and UV lines within these same
warm and diffuse local gas clouds. Furthermore, UV observations of stars within the
$\beta$ CMa interstellar tunnel suggest that although the neutral gas density along
this 200pc long interstellar feature may be very low, this is not the case for
ionized gas components (Dupin $\&$ Gry \cite{dupin98}). Clearly it would
be instructive to compare maps of the spatial
distribution of CaII absorption (which is sensitive to partially ionized gas) and
those derived from NaI observations (which trace cold and neutral gas) for the local ISM.

In this Paper we report on the results of an extended survey of NaI and CaII absorption lines recorded
at high spectral resolution (R $>$ 50,000) towards many newly observed early-type stars mostly located
within 800pc of the Sun. We present new absorption measurements of the NaI doublet recorded
along 482 sight-lines, in addition to 807 new measurements of CaII absorption along 807 sight-lines.
When added to similar NaI and CaII absorption measurements reported in the
literature, we are now able to present new maps to 300pc of the 3-D distribution of neutral and
partially ionized gas towards 1678 stars (at NaI) and 1267 stars (at CaII). Although these new data
confirm many of the findings of Paper 1, they also show several new interstellar
gas features now
revealed through the increased level of sampling of  local ISM sight-lines, especially
at high galactic latitudes and in regions with
distances in the 150 - 350pc regime. The 3-D maps of 
local CaII absorption are the first of their kind to be presented, and in several respects they mimic many of
the interstellar features seen in the equivalent maps of local NaI absorption. However, there are
a significant number of regions where CaII absorption is high and NaI absorption is low, and vice-versa.
Such variations in the observed column density ratio of N(NaI)/N(CaII)  are influenced by the conditions in the local stellar
ionization field. Values of this ratio for stars near to the galactic plane in each of the 4 galactic
quadrants are also presented.
Finally, as a by-product to determining the (NaI and CaII) gas column densities
along each sight-line we have also determined the absorption velocity (i.e. cloud component) structure towards
each target. These results will be published in a separate Paper that deals with the kinematics
of local gas clouds within $\sim$ 150pc (Lallement et al. \cite{lall10}).

\section{Observations and Data Reduction}
We present observations of both the interstellar NaI D-line doublet at $\sim$ 5890\AA\ and
the CaII K-line at 3393\AA\  recorded in the absorption spectra of early-type stars
with distances $<$ 0.8 kpc.
These data were obtained during observing runs performed
over the period 2003 - 2008 using the following instrumentation (i) the Aurelie spectrograph
at the 1.52 m telescope of the Observatoire de Haute Provence (France), (ii) the Hamilton
echelle spectrograph on the 0.9m coude feed telescope at the Lick Observatory (USA),
(iii) the Hercules echelle spectrograph on the 1.0m McLellan telescope at the Mt. John
Observatory (New Zealand), (iv) the FEROS coude echelle spectrograph on the 2.2m telescope
at the European Southern Observatory (Chile) and (v) the GIRAFFE fiber-fed echelle spectrograph
at the 1.9m Radcliffe telescope of the South African Astronomical Observatory (RSA).
The spectral resolution for these interstellar data was 3 km s$^{-1}$ for the Observatoire de Haute Provence (OHP) data, 5 km s$^{-1}$ for the Lick Observatory data, 4 km s$^{-1}$ for the Mt. John
Observatory (MJO) data, 6 km s$^{-1}$ for the  FEROS ESO data and 7.5 km s$^{-1}$ for the South African
Astronomical Observatory (SAAO) data. 

Interstellar sight-lines were generally selected using the criteria that the background stellar targets possess:
(i) a Hipparcos catalog distance (ESA \cite{esa97}) normally less than 400pc and with an associated relative standard
parallax error
smaller than 0.3 (however several targets with distances up to 800pc were also included 
for certain sight-lines of specific interest), (ii) a spectral type earlier than A5V, (iii) a stellar rotational velocity $>$ 25 km s$^{-1}$,
(iv) a galactic position that filled in areas of the sky not well sampled by the targets listed
in Paper 1, and (v) a sufficiently bright visual magnitude that a well-exposed spectrum could be gained within
a 60 minute exposure.
We also avoided observations of known spectroscopic binary stars, since interstellar features
are often difficult to identify within their stellar absorption profiles. In Table 1, available on the CDS, we list information on the newly observed
targets that includes their HD number,  galactic longitude and latitude, Hipparcos distance (in pc)
and the observatory (OHP, Lick, MJO, ESO and SAAO) where the data was taken, together with the year of
observation. 

All of the data, except for the MJO and ESO observations, were reduced using software
routines outlined in detail in Sfeir et al. (\cite{sfeir99}). These routines involve:  (i) division of
the raw data by a flat-field,
(ii) removal of instrumental background light and
cosmic rays, (iii) removal of telluric water vapor lines (which particularly affect
the NaI D2-line at $\lambda$ 5890\AA\ ) using either
a synthetic telluric transmission spectrum as detailed
in Lallement et al. (\cite{lall93}) or by division by a purely
stellar-atmospheric absorption spectrum of an unreddened (nearby)
B star observed at a similar altitude and time, and (iv) spectral order
extraction and wavelength calibration using Th-Ar lamp spectra recorded
several times throughout each observing night. Similar data reduction processes were performed on the
Mt. John Observatory and ESO data using their own in-house standardized
echelle data extraction software packages.

The wavelength calibrated spectral data were subsequently fit with a high order polynomial in order to
define a local stellar continuum placement, such that the equivalent widths (W$_{\lambda}$)
of the NaI and CaII
absorption lines could be measured from these resultant intensity profiles. The errors on the
measured values of W$_{\lambda}$ are dominated by the statistical uncertainty on the measured
counts within each line profile and the systematic uncertainty in the local
continuum placement (see Welsh et al. \cite{welsh90} for a fuller discussion).
These two sources of error are added in quadrature to derive the measurement error.
Typically, for a well-exposed spectrum, a measurement error of $\sim$ 5$\%$ is normal for
our data.
Upper limits to the value of equivalent width where no
absorption line was detected with significance were derived from conservative estimates of
the strength of a potential absorption feature appearing at a level $>$ 2.5-$\sigma$
above the rms value of the local continuum. For the majority of our well-exposed spectra
(i.e. S/N $>$ 30:1) this typically resulted upper limit values of W$_{\lambda}$ $<$ 2.5m\AA\ for
both the NaI-D2 and CaII-K lines. The measured values of W$_{\lambda}$ for
the NaI (D2 $\&$ D1) and the CaII-K lines are listed in Table 1 for our presently observed sight-lines.

For the majority of targets with distances $\la$ 300pc both the interstellar NaI and CaII absorption
lines are not greatly saturated. Thus, we were able to accurately fit these profiles with absorption
components (i.e. gas clouds) using the line-fitting procedure described in Sfeir et al. (\cite{sfeir99}).
A best-fit theoretical absorption profile is generated that is characterized by 3 parameters:
(i) an interstellar cloud (heliocentric) velocity, V (ii) a doppler velocity dispersion parameter
($\it b$-value) and (iii) a cloud component column density, N(NaI) or N(CaII). The best-fit models
were further constrained by fitting $\it both$ of the NaI D-lines simultaneously (with the fit
being biased towards the stronger D2 line). For the ESO data, in which both the CaII K and H-lines
were available, the best-fit for the doublet was similarly constrained.
 
As an indication of the quality of the data recorded
with the spectrograph+telescope systems at each of the 5 observatories, in Figures 1- 5 we
show some typical model fits 
to the observed interstellar NaI and CaII absorption profiles
recorded towards stars with distances in the 50pc - 500pc range. The best-fit model
parameters (i.e. V, $\it b$ and N(NaI)) and/or N(CaII) for the (several) cloud components required
to fit each of these absorption profiles are listed in a box beneath each of
the profiles in these Figures. The fits were performed using the minimum number of
absorption components, with the addition of extra components only being deemed
necessary until the residual between the model fit and the data points was
approximately equal to the rms error of the data.  In general (dependent on the spectrograph used), the accuracy
of the placement of each of the cloud component velocities within these line-fits is $\sim$ $\pm$0.5 km s$^{-1}$.

The highest resolution data (see Figure 1) was obtained for the CaII K-line
at the OHP, and hence it is no surprise that these data generally require more absorption components
to fit the observed profiles. In contrast, the data gained with the lower resolution FEROS system on the 2.2m telescope
(shown in Figure 4) is of the highest S/N ratio and allows access to both the NaI and CaII doublets, such that the model fits
can generally be constrained to a better degree of confidence.
However,
for the purposes of our present interstellar study we only require use of the total
equivalent width, W$_{\lambda}$(D2) and W$_{\lambda}$(K),  and the
summed total column density values, N(NaI)$_{tot}$ and N(CaII)$_{tot}$, measured towards
each star. These values are listed in Table 1.  The best-fit values of NaI and CaII column density
for each cloud component for all of the newly observed targets will be presented in a forthcoming paper which will
concentrate of the velocity structure of the local ISM within 150pc (Lallement et al \cite{lall10}).

In order to increase the spatial sampling of interstellar gas absorption
in the local ISM, in addition to the newly obtained data listed in Table 1 we 
have also
included NaI and CaII interstellar measurements
(of equivalent width and associated total column density) towards sight-lines $<$ 800pc that have been
previously published in the literature. The corresponding data for each of these
observations are also listed in Table 1 along with
the appropriate reference for the particular observation. For cases in which only values of equivalent
width are quoted in the literature, we have derived a corresponding value of total column density (reported
in Table 1) using
empirical curves of growth for both NaI  and CaII derived from all of the other data. We have
not included interstellar data for targets with known (or suspected) circumstellar gas, since this can
provide a significant (and variable) contribution to the values of NaI and
CaII column density (Welsh et al. \cite{welsh98}).
One important case of this type
is the star HD 225132 (2 Cet), which appeared in the NaI maps
of Paper 1 as a nearby (d = 70pc) condensation of cold gas lying within the confines of the LC. Recent high
resolution measurements of NaI  and CaII absorption by Redfield et al. (\cite{red07}), which have subsequently
been confirmed by our own measurements, revealed a far smaller value of NaI column density for
this sight-line compared to that reported in Welsh et al. (\cite{welsh94}). In addition, our own CaII measurements
reveal a component at V = +20 km s$^{-1}$ not seen by Redfield et al. We have therefore removed this star
from our compilation in Table 1.
In total
we now present 1678 NaI and 1267 CaII interstellar sight-line absorption measurements, 1088 of which have 
common sight-line measurements for both of these ions.

In Figures 6(a) and 6(b) we show the galactic distribution of the respective 1678 NaI
and 1267 CaII sight-lines sampled. Although the sky-coverage is significantly increased since
Paper 1, we note some areas of the sky (e.g. $\it l$ = 295$^{\circ}$, b = -45$^{\circ}$ and 
$\it l$ = 160$^{\circ}$, $\it b$ = -75$^{\circ}$) remain sparsely sampled. This
is primarily due to the paucity of suitably bright O, B and A-type stellar targets in these
directions that can be
observed in a reasonable exposure time with the presently available instrumentation.
Therefore any conclusions derived from the mappings of interstellar absorption
towards both of these galactic directions must be viewed with some caution. We also
note that our selection of suitable targets is biased against heavily reddened and
visually faint (m$_{v}$ $\ga$ 6.5) sight-lines, due to the practical restriction of $\sim$ one hour of exposure time
per recorded spectrum.

\section{Interstellar Absorption Mapping}
The measurements of NaI and CaII equivalent width and column density measured along the 1857 sight-lines allows
us to probe the absorption characteristics of both neutral and partially ionized local interstellar gas. Our underlying assumption
is that the NaI ion is a good tracer of of the total amount of neutral and cold (T $<$ 1000K) interstellar gas along most galactic
sight-lines (Hobbs \cite{hobbs78}). Similarly, the CaII ion is assumed to trace both neutral and warmer (T $\le$ 10,000K)
partially ionized gas clouds. Previous high resolution studies of NaI and CaII absorption suggest that the latter ion has larger
absorption line-widths for components seen in NaI absorption at
the same velocity (Welty, Morton $\&$ Hobbs \cite{welty96}).
This has been interpreted as the CaII  gas components occupying a larger volume than
those of NaI, with an associated higher temperature and/or more turbulent velocity than the volume of gas occupied by NaI.
In our present study, in which we average the absorption properties of gas along many interstellar sight-lines, we assume
that the derived spatial distributions of both NaI and CaII are directly comparable. We note that CaII absorption
not only arises in cold (and relatively dense) gas in which Ca may be heavily depleted onto dust grains and is the dominant
ionization state, but it can also be present in
warmer and lower density gas clouds where Ca is less depleted and in which CaII is a trace ionization state.

Our primary aim in this paper is to determine the 3-D spatial distribution of 
interstellar gas (as characterized by NaI and CaII absorption) to a distance of $\sim$ 300pc from the Sun.
Following our previous work on this subject (Sfeir et al \cite{sfeir99}; Lallement et al. \cite{lall03}), we produce
plots of the spatial distribution of gas based on the inversion of
column densities using a method originally devised by Vergely et al (\cite{verg01}). This 
method of plotting, together with some of its advantages and shortcomings, has
been described fully in Lallement et al. (\cite{lall03}). However, before presenting such gas density maps we
firstly discuss the measurements of the equivalent widths of the interstellar NaI
and CaII lines.

\subsection{NaI and CaII Equivalent Width Values}
In Figure 7 we show a plot of distance versus the total equivalent width of
the NaI D2-line, W$_{\lambda}$(D2), for targets with distances $<$ 400pc. Note that
this plot does not include sight-lines that have reported values of N(NaI), but no associated
information on the measured equivalent width.
This figure, in agreement with a similar plot shown in Sfeir et al. (\cite{sfeir99}), shows
that very little measurable NaI absorption can be detected for distances (in all galactic directions)
up to $\sim$ 80pc from the Sun.
This general lack of appreciable NaI absorption (W$_{\lambda}$(D2) $\la$ 5 m\AA)
provides clear evidence that the volume of the
LC is essentially free of major condensations
of cold and dense neutral interstellar gas.
Beyond $\sim$ 80pc the level of NaI absorption 
rises sharply over a short distance of $\sim$ 20pc to reveal
the presence of a dense `wall' of neutral gas (W$_{\lambda}$(D2) $>$ 20 m\AA) that surrounds the LC in many galactic directions
(Welsh et al. \cite{welsh94}). Beyond the LC neutral boundary, which is normally characterized by a single narrow velocity
component in the NaI absorption profiles, more and more NaI absorbing (cold) neutral interstellar clouds are encountered as sight-line
distances increase. In several galactic directions with distances $>$ 300pc the NaI D-lines become increasingly
saturated and we therefore generally see a slower
increase in equivalent width value with greater distance sampled.
However, we also note the presence of  several sight-lines with distances $>$ 250pc that have minimal associated
NaI absorption (plotted as open circles in Figure 7). These are sight-lines which extend into the lower galactic halo
through the openings of the Local Chimney, a region of known low neutral gas density (Crawford et al \cite{craw02}). 
The nearest stars with distances
less than 65pc, but with `anomalously high'  values of W$_{\lambda}$(D2) $>$ 10 m\AA\ , are
HD 184006 (d = 38pc), HD 159170 (d = 48pc),
HD 186882 (d = 52pc), HD 105850 (d = 56pc), HD 96819 (d = 58pc) and HD 129685 (d =63pc).

In contrast with studies of interstellar NaI, previous investigations of CaII absorption have
shown that the local ISM is composed of many diffuse and
partially ionized  warm (T $\sim$ 7000K) cloudlets that
possess a complex velocity structure, even over distances
as short as 5pc (Crawford, Lallement $\&$ Welsh \cite{craw98}, Lallement et al. \cite{lall86}).
These cloudlets are randomly scattered within 50pc of the Sun and although are generally not revealed
by NaI absorption measurements, they have been extensively studied at ultraviolet
wavelengths (Redfield $\&$ Linsky \cite{redfield08}).
In Figure 8 we show that the behavior of CaII absorption as a function of distance (to 400pc)
does not  follow the same pattern that
was found in Figure 7 for NaI measurements. For the majority of sight-lines within $\sim$ 100pc
we mostly find W$_{\lambda}$(CaII-K) $<$ 15 m\AA, with no sharply increasing
value of absorption at $\sim$ 80pc that might feasibly
be associated with the presence of gas associated with
the boundary wall to the LC. Instead there seems to be a wide-spread local distribution of 
partially ionized CaII regions with a  range of low integrated CaII absorption strengths that extends
to at least 100pc.  Beyond
 $\sim$ 100pc  we observe a slowly increasing value of
W$_{\lambda}$(CaII-K)  with the sight-line distance sampled. Even by 200pc there are
still several sight-lines in which very little ($<$ 3m\AA) interstellar CaII is present. 
At a distance of $\sim$ 300pc  a value of  W$_{\lambda}$(CaII-K) $<$ 100m\AA\ is 
typical for
most of the galactic directions sampled, although values in excess of 200m\AA\ are found
for a few higher density regions.

Welsh et al. (\cite{welsh97}) have presented plots of the total column density of CaII versus the total column density
of neutral HI which show that both quantities are broadly correlated. However, there is significant
scatter in this relationship due to the fact that
along any given interstellar line-of-sight the measured amount of CaII-bearing gas is closely
linked to the ambient ionization conditions, the amount of dust depletion and the size and number of gas clouds
encountered. 
The Local Cavity is an anomalous region of 
interstellar space that contains many randomly scattered warm and partially ionized low
column density CaII-bearing
cloudlets (each of W$_{\lambda}$(CaII-K) $\sim$ 5 m\AA, Redfield $\&$ Linsky \cite{redfield02}), probably
mostly of similar physical size. Based on our fitting of CaII absorption profiles for stars
$<$ 100pc, the number of `cloudlets' encountered along a typical interstellar sight-line is generally $<$ 4.
Thus, the presently measured total equivalent width of CaII along sight-lines
$<$ 100pc is small ($<$ 15 m\AA) and variable, dependent
on the number of warm clouds encountered. For cold interstellar clouds that
exist beyond the rarefied Local Cavity,  it is generally assumed that the warm CaII 
gas is present in the `outer-skins' of these clouds
(Vallerga et al. \cite{vallerga93}, Welty et al \cite{welty96} ). Thus, although
the neutral wall boundary to the LC may contain significant amounts of cold NaI gas compared
with sight-lines lying within the LC, the corresponding increase in the equivalent width of CaII due
to the outer skin of this boundary wall is far less pronounced. Hence, we do not
observe a sharply increasing
value in CaII absorption at $\sim$ 80pc due to 
the presence of the boundary wall of the LC. 
For greater sight-line distances in
the general ISM (i.e. d$>>$ 100pc), the measured CaII equivalent width 
should be more closely linked to the
number of clouds (or cloud envelopes) sampled along a particular sight-line as opposed to the actual distance sampled. 
Thus we expect to observe a gradual (but variable) increase in CaII absorption equivalent width with distance, as
is demonstrated in the increase of N(CaII) with N(HI) by Welsh et al (\cite{welsh97}).

We note that there are only 3 sight-lines within 75pc
with anomalously large values of W$_{\lambda}$(CaII-K)  in the 20 - 30m\AA\  range. These
are the sight-lines towards the stars $\alpha$ Oph (d = 14pc), HD 145570 (d = 51pc),
and HD 113852 (d =70pc). Also, the sight-lines towards HD 5487 (d = 259pc), HD 85504 (d = 322pc) and
 HD 75855 (d = 446pc) have anomalously low
levels of CaII K-line absorption (i.e. log N(CaII) $<$ 10.5 cm$^{-2}$).  These
are all high latitude sight-lines that extend through the
openings of the low density Local Chimney into the overlying galactic halo.


\subsection{The Volume Density Distribution of NaI and CaII}
Before progressing to a discussion concerning the creation of 3-D density maps of the spatial distribution of both
interstellar NaI and CaII gas density,
it is informative to firstly investigate the global variation of volume density of both ions as
a function of distance sampled.
In Figure 9 we show the plot of NaI volume density, n$_{NaI}$, (i.e. N(NaI)/distance) versus distance (pc), with upper limit
values being plotted as open circles and measured values plotted as filled circles. We note that the lowest values
of n$_{NaI}$ are mostly found for distances $<$ 200pc (i.e. the region of the ISM dominated by
the effect of the low density LC), although there are several sight lines (mostly at high galactic latitudes) with distances up
to 600pc that have similarly small volume density values. In the 110 - 200pc distance range there
are about a dozen stars with slightly higher than average
values of n$_{NaI}$ 
in the range 10$^{-7}$ - 10$^{-8}$
cm$^{-3}$. The majority of these stars lie in the longitude range $\it l$ = 350$^{\circ}$ - 50$^{\circ}$, in
the general
direction of the galactic center.
For sight-lines $>$ 300pc we see a general trend in the data such that 
values of n$_{NaI}$ in the 10$^{-8}$ to 10$^{-10}$ cm$^{-3}$ range are representative
for most distances up to 800pc. We interpret these latter data points as indicating that
the general ISM has a wide range of NaI volume density values, each of which is critically dependent on the number
and density of neutral interstellar clouds and interstellar cavities encountered over any given sight-line distance.

In Figure 10 we plot the volume density of CaII, n$_{CaII}$, as a function of distance, whose
values for sight-lines $<$ 100pc show a complex pattern of behavior. For distances $<$ 30pc we measure
both very high and very low values of n$_{CaII}$, dependent on whether a particular sight-line encounters
any of the local fluff clouds. For distances in the 30 - 100pc range, although there are still sight-lines
with very low values of n$_{CaII}$ $<$ 10$^{-10}$cm$^{-3}$,  a value of $\sim$ 5 x 10$^{-10}$cm$^{-3}$ seems fairly
representative for most interstellar directions. However, for distances $>$ 100pc higher values
of n$_{CaII}$ are generally encountered in the ISM,
such that a volume density of $\sim$10$^{-9}$cm$^{-3}$
is typical for most sight-lines with distances up to 800pc. This would suggest that beyond $\sim$ 100pc
the density distribution of CaII bearing gas is fairly uniform throughout the ISM. This is in accord
with the findings of Welsh et al. (\cite{welsh97}) and Megier et al. (\cite{megier05}) who both found
that the level of interstellar CaII absorption is far better correlated with distance
for sight-lines  $>$ 100pc than for NaI absorption.

\subsection{Column Density Inversion Maps}
Although the plots of equivalent width versus distance can often reveal how absorption 
accumulates over sight-lines (by revealing gradients
in the gas density distribution), the significance of the highest levels of absorption is lost at large distances due to line saturation, and similarly
information is lost on absorption with levels less than that of the lowest measurable limit of
equivalent width. However the column density inversion method of Vergely et al. (\cite{verg01}) used in 
Paper 1,
which derives volume gas densities in 3-D space from the integrated sight-line column density values by
an iterative inversion process, can reveal masses of
gas at any distance provided there are sufficient constraining target measurements.

The inversion algorithm we have presently used to construct the 3-D gas density maps is derived from a
non linear least-squares approach to the generalised inverse
problem (Tarantola $\&$ Valette \cite{taran82}). 
It differs from a maximum likelihood method in that it allows treatment
of under-determined cases by imposing some additional
constraints. For the present NaI and CaII data we use `a priori' information on 
the gas density that
follows a Gaussian law with non null spatial autocorrelation.
In other words, the computed gas density
distribution can be seen as fluctuations around this `a priori' solution.
The mathematical specificity is that the solution belongs to the L2 Hilbert
functional space, which allows retrieval of the solution everywhere without
`a priori' space discretization.
For the case of a linear operator linking the unknown parameters
(the gas densities) and the data (the measured columns) and an assumed
Gaussian fluctuation around the `a priori' value, a solution can be
determined directly in one iteration. However this is not the case for
our present data (i.e. column densities and distances), since
in order to force the gas density $\rho$ to be positive we use a change of variable, namely $\alpha$ = log($\rho$ / $\rho$$_{0}$) , $\rho$$_{0}$ being a constant.
In this case the solution to obtaining the gas density as
a function of distance is through an iterative
method, as described in Tarantola $\&$ Valette (\cite{taran82}) and Vergely et al
(\cite{verg01}).

The inversion method was first applied to opacities in 3-D space by
Vergely et al. (\cite{verg01}). Here we use the same algorithm, but with the
following constraints that are specific to our present set of data: (i)
the correlation between the gas density at point X and the density at point
X' is of the form exp(-$\mid$X - X'$\mid$$^{2}$ / $\xi$$^{2}$), with a correlation length, $\xi$,  of 20 pc.
This distance corresponds to the maximal resolution allowed by the present
data set and is roughly the average distance between two target stars.
This is an improvement upon our previous inversion of NaI data which used
$\xi$ = 25 pc, the  improvement  being allowed by the present larger data set. It implies that
details smaller than this characteristic scale will be smoothed out and
appear at least 20 pc wide. It also means that the computed volume
densities are not the local values, but are mean values in volumes whose size
is of the order of 20$^{3}$ pc, 
(ii) the `a priori' density decreases exponentially with the distance from the
galactic plane, with a scale height of 170 pc for Na and 450 pc for CaII.
These two values were determined from the data sets themselves by fitting
an exponential law (for the density) to the measured columns.
(iii) the `a priori' error on the opacity, $\alpha$ = log($\rho$ / $\rho$$_{0}$), follows a Gaussian
law with a standard deviation $\sigma$($\alpha$) = 2.5. This allows high
contrasts for the density, i.e. very dense clouds or close to null values.
(iv) the errors on the data follow a particular law chosen to
minimise the effects of potential outliers within the data set, as originally
devised by Vergely et al. (\cite{verg98}), namely : 


\begin{equation}
f(c(i)-c_{0}(i))  = \pi \cdot \sigma(i) \cdot cosh( \frac{(c(i)-c_{0}(i)}{\sigma(i)} )
\end{equation}

where f(c(i)-c$_{0}$(i)) is 
the probability for a specific column density c(i) towards target i  to depart from
the central data value c$_{0}$(i). 
Developments show that for close values this function tends to a classical Gaussian (a L2 norm),
while for values far above or below the central one it tends to vary like
exp(-(c(i)-c$_{0}$(i))/$\sigma$(i)), i.e. a L1 norm.

(v) Relative errors on the columns are taken to be 25$\%$ for all targets, and those errors are combined with the individual errors on the Hipparcos parallaxes.

Finally, we want to emphasize that within our data set of gas densities there are some locations for which there is no
information that can change the volume density from the initial `a priori'
value. For these locations the gas density value is kept unchanged. We have inluded such information in our maps (see below).

The resultant accuracy of the inversion method has been evaluated by Vergely (\cite{vergely98}) through extended tests using synthetic cloud distributions. The comparison between these initial synthetic distributions and the resultant inverted densities shows that there are optimal values for both the correlation length (also called the smoothing parameter, x) and the statistical error on the density (i.e. the allowed density fluctuation amplitudes). The correlation length is clearly linked to the average distance between the target stars and the optimal selected value essentially defines the resulting placement accuracy of gas clouds using the inversion method. In our case this distance is of the order of 35 parsecs for the entire target sample, but is less at smaller distances from the Sun. This is because nearby targets are more numerous, and thus more  information can be obtained within the first 150 parsecs. Hence we have selected the smaller value of 20 pc for the correlation length in our new maps. However, the standard deviation for density fluctuations must be non-negligible, since we know in advance that there are dense areas and voids between the many groups of dense clouds. 

In order for the reader to appreciate the typical accuracy of the inversion method for the
placement of interstellar gas clouds,  in Figure 11 we show
an example of a fractal distribution of measurements of
interstellar reddening in the galactic plane and the corresponding inversion of cloud densities computed for 2500 sight-lines randomly distributed in distance and directions around the galactic plane. These data are taken from
Figures 5.10 and 5.11 of Vergely (\cite{vergely98}), with 
a correlation length of 15 pc and measurement errors being simulated as `noiseÕ in the reddening and distance data. Vergely (1998) found that the broad structures of reddening (which are equivalent to gas clouds) were well located after the inversion, but clouds (or voids) smaller than 10-20 pc were included in broader cloud structures and thus smeared out. Finally, it must also be mentioned that since the inversion method is statistical in nature, it is impossible to prevent some bad restitution for one or more cloud structures. This may happen when the combination of the actual distribution in one particular galactic area and the target sampling in that area are highly non-uniform (i.e. when the region or galactic direction is sparsely sampled). 

The end-product of the inversion method, when applied to our present
measurements (and associated errors) of column density and distance for each of the sight-lines sampled,
is a 3-D data cube of NaI (or CaII) gas density as a function of  distance and galactic longitude and latitude.
Our new NaI volume density data cube
has a 60$\%$ increase in the number of targets over that of Paper I, and the CaII maps are the first ones of their type ever to be presented. In Paper 1 the correlation length for the NaI data was $\sim$ 25pc, whereas
the gas distribution shown in our new maps of NaI and CaII
can be considered representative at the 20pc 
sampling scale.
Thus, as mentioned above, small-scale absorption features (i.e. small cloudlets of low density gas) could easily be missed in our final smoothed density maps. For a few cases very high column density
values of log N(NaI) and log N(CaII) $>>$ 13.5 cm$^{-2}$ are reported by other authors in Table 1. 
In some cases it was evident that the absorption profiles had not been fitted with sufficient cloud components
and we noted
that such very high density values tended to bias and unrealistically distort the 3-D mapping program (with associated points 
of inflection)
when low density sight-lines are spatially close to these apparently very dense regions. In order to smooth out this effect, but still
retain the presence of the high density regions in our maps, we determined that by generally limiting the maximum values
of log N(NaI) and log N(CaII) to 13.5 cm$^{-2}$ this distortion effect was minimized in the resultant plots.

Since we have created 3-D data cubes for both NaI and CaII
gas density out to a distance of 300pc from the Sun, in principal we can present maps of the spatial density
distribution of either ion in any desired galactic projection. However, for brevity, we present maps for the 3 main
projections (galactic, meridian and rotational planes) for both NaI and CaII ions. It is
envisaged that the entire data-base of
absorption measurements, together with the associated 3-D gas density cubes for NaI and CaII  will be made available
to the community on the world-wide web by the end of 2009.

\section{NaI absorption within 300pc}
In the following sub-sections we present maps of the 3-D distribution of
NaI absorption within 300pc of the Sun, formed from three main galactic projections through the 
previously described
3-D NaI gas density cube. Each of the maps represents
the spatial distribution of the volume density of NaI (n$_{NaI}$) as a function of distance and
thus is a 3-D visualization of the data previously discussed in Section 3.2.
Each map shows the regions where
very low values of neutral gas density are located (white shading) together with regions
where statistically significant amounts of neutral gas have been detected (grey/black shading).
Areas with insufficient information to perform an inversion of the columns are indicated by a
matrix of dots in the maps. They
correspond to areas within which the density is still the `a priori'
density. In cases where some information is present but the number of constraints
from the target stars makes the location of the cloud, or of the cavity,
very imprecise, the inversion process may create `finger-like' extensions
that are easily recognisable in the maps. The presence of such features can be interpreted as
indications of the existence of a gas cloud along the `finger' direction,
but the location of the accumulation of gas is thus uncertain.

Stars used to create each map are shown as triangles with their size being proportional to their distance above
(triangle vertex pointing up) or below (vertex pointing down) the galactic plane. 
Superposed on the maps are contours of iso volume density with values of log n$_{NaI}$ = -9.5,
-9.1, -8.5 and -7.8 cm$^{-3}$ shown respectively in yellow, green, turquoise and blue lines.
As an approximate guide, these average values of neutral gas volume density can be converted into
hydrogen density values at a particular distance using the relationship given in Ferlet et al. (\cite{ferlet85}).
Finally, we further remind the reader not to over-interpret every absorption feature that
appears in these maps due to the reasons listed previously. 
 
\subsection{NaI: A view from above the galactic plane}
The greatest number of sources (i.e. early-type stars) sampled by this survey lie
close to the galactic plane and hence the gas density maps derived
for both NaI and CaII for this particular galactic projection are the most accurate.
In Figure 12 we show the spatial distribution of NaI  density to 300pc which clearly reveals a
large and highly irregularly shaped volume of space surrounding the Sun 
to a distance of $\sim$ 80pc in most directions that is devoid of dense condensations of cold
and neutral gas clouds. This region, of course, can be identified as the Local Cavity (LC). 
We note that the LC is more elongated in galactic quadrants 3 and 4 compared with the other two
quadrants, with quadrant 1 having the smallest volume of low-density neutral gas.
Our new map also reveals for the first time that the LC (in the galactic plane) can
essentially be considered as two
 rarefied `sub-cavities' that abut each other along a thin interstellar gas filament of
low column density ((log N(NaI) $\sim$ 10.4 cm$^{-2}$) at a distance of $\sim$15pc in the direction of
$\it l$ $\sim$ 345$^{\circ}$. One of these `sub-cavities' is mostly contained within
galactic quadrants 2 and 3 (and to a far lesser extent within quadrant 1), with the majority of the other low density `sub-cavity' mostly being contained within
quadrant 4. The appearance of smaller `cell-like' cavity structures bounded by thin gas filaments 
is also hinted at in the equivalent maps of
CaII absorption and these will be discussed further in Section 5.1.

The relatively dense neutral gas wall that surrounds the LC in the galactic plane
has a typical depth of
50 -80pc in most directions, but  
it is quite evident from the map that the LC void is $\it not$ fully enclosed
by a continuous boundary of cold and neutral gas. Instead, the LC can be
best described as being `porous' in the sense that it appears to be linked to several adjacent
interstellar cavities through narrow gaps in the surrounding neutral gas wall. 
These adjacent interstellar cavities include
the Loop I superbubble, the $\beta$ CMa interstellar tunnel and the Pleiades bubble, all of which
were revealed in the preliminary maps presented in Paper 1. 
Our new map now reveals that the tunnel of low interstellar gas density towards the direction
of the star $\beta$ CMa ( $\it l$ = 226$^{\circ}$) is bifurcated at a distance of $\sim$160pc, which is also
the distance to $\beta$ CMa itself. The tunnel's most pronounced region of low neutral
density extends to at least 250pc in the direction of  $\it l$ $\sim$ 260$^{\circ}$, which provides a pathway
to the GSH 238+00+09 supershell which may have been created
by the star cluster Collinder 121 (Heiles \cite{heiles98}).
The pathway to the  Loop I region ($\it l$ $\sim$ 345$^{\circ}$) is now revealed
to be far narrower than that originally shown in Paper 1, with only one low density passageway leading
towards the Sco-Cen OB association. The present map shows this narrow entrance to consist
of several fragmented clouds located over the
distance range 90 - 120pc, in agreement with previous UV observations of this region by Welsh
$\&$ Lallement (\cite{welsh05}).
Our new map shows three other regions of low neutral gas density (interstellar bubbles?)
that exist  beyond the neutral boundary to the LC at distances $>$ 200pc. These are located in the
directions of
$\it l$ = 190$^{\circ}$ (Taurus), 210$^{\circ}$ (Orion/Eridanus) and 285$^{\circ}$ (Carina).
 
Figure 12 also reveals a previously unknown
gap in the neutral boundary to the LC in the galactic plane that lies between $\it l$ = 70$^{\circ}$ to 80$^{\circ}$. 
This low density extension of the LC  lies in the direction of Cygnus and
may be linked to part of the Loop II SNR which lies at a distance of $\sim$ 110pc  (Berkhuijsen \cite{berk73}).
We note that in the direction of this newly discovered gap in the LC wall
there is a large patch of emission revealed in both
the 3/4 keV and 1/4 keV maps of the SXRB (Snowden et al \cite{snow97}).
The presence of such (low neutral gas density) gaps and openings in the wall of the LC should have a profound effect on the
potential observability of more distant sources of soft X-ray diffuse emission. Although this effect
is most pronounced when viewing the halo through
the openings of the Local Chimney at high galactic latitudes (see Section 4.2),
the gaps may also provide a low absorption path for emission from soft X-rays generated in nearby
hot superbubbles. For example it is well-known that there is a dipole effect in
the distribution of the soft X-ray background emission, with the emission being
warmer, of greater intensity and with a larger spatial coverage in directions towards the
galactic center than along sight-lines towards the galactic anti-center at mid-plane latitudes
(Snowden et al \cite{snow00}). This effect may now be explained through inspection
of Figure 12 which shows that the local void is linked to the 
Loop I superbubble (which
is a known source of million degree soft X-ray emission)
through a gap in the surrounding boundary wall
in the direction of the galactic center.

The closest accumulations of neutral gas in the galactic plane (with log N(NaI) $\sim$ 10.8 cm$^{-2}$) 
are located in the directions of (i)
$\it l$ $\sim$ 85$^{\circ}$ at a distance of $\sim$ 38pc towards the star HD 184006,
(ii) in the general direction of the galactic center at a distance of $\sim$ 50pc (towards HD 159170),
and (iii)
towards several stars with distances in the 60 - 80pc distance range
in the direction of $\it l$ $\sim$285$^{\circ}$. All 3 of these low neutral
density gas clouds are extensions of the surrounding denser neutral wall that
protrude into the central regions of the LC. They appear to form the filamentary gas boundaries to the
inner cell structures whose possible presence within the larger LC was
alluded to previously.

In addition, in the direction of $\it l$ $\sim$195 $^{\circ}$ at a distance of $\sim$ 95pc
we detect another small, more dense gas cloud.  The  presence of two small gaps in the
 neutral wall to the LC in this general direction at $\it l$ = 165$^{\circ}$ and $\it l$ = 190$^{\circ}$
suggest that the isolated dense cloud  is most probably part of the broken boundary wall to the LC. 
Finally, we note that although our sampling of targets with distances in the 200 - 400pc range
is far from complete (as indicated by regions of small dots in the map), there
do appear to be galactic directions in the galactic plane in which the neutral gas
density does not increase with the distance sampled. For example, in the direction of $\it l$ = 315$^{\circ}$ 
we have an appreciable number of targets sampled with distances $>$ 200pc which show no additional
NaI absorption beyond that distance. This suggests that there may be other large regions of low neutral gas
density (i.e. superbubbles) that remain to be revealed by future more extensive absorption measurements 
in the interstellar region between 200 and 400pc from the Sun.

\subsection{NaI absorption viewed in the meridian plane}
In Figure 13 we show the spatial distribution of NaI absorption to 300pc as projected
in the meridian plane with the galaxy essentially being viewed in a side-on manner. The plotting
symbols are the same as those for Figure 12.
The cold and dense NaI absorption associated with gas in the galactic plane can be seen
as a $\sim$200pc (vertically) thick gaseous bar that stretches from left to right across Figure 13. 
The low neutral density
void in the central region of this plot is the LC, which in this projection can be identified with
the open-ended Local Chimney that is tilted at an angle of $\sim$ 35$^{\circ}$ from
vertical. This new plot confirms the recent studies of Crawford et al. (\cite{craw02})
and Welsh et al. (\cite{welsh04}), in which no continuous neutral boundary
to the LC can be found at high latitudes in either galactic hemisphere.
This map also confirms the general lack of neutral gas for sight-lines
that extend to at least $\sim$175pc into the inner halo regions. We also note
that the rarefied Local Chimney region fragments into 
finger-like extensions that reach into the inner halo in both hemispheres.  Filaments
of neutral HI gas have been widely observed in the general ISM and their
appearance within the local ISM suggests that they
may be linked to the origin of the LC in which an explosive event  may have 
`cleared out' any neutral gas that may have once resided close to the Sun and
ejected it into the overlying galactic halo.

Figure 13 shows four major condensations of cold neutral gas
lying $\it within$ the confines of the LC in this galactic projection.
They appear to lie along a fragmented bar of dense gas that
lies at an angle of $\sim$45$^{\circ}$ to the galactic plane that stretches
from 200pc below to 30pc above the plane. The nearest of these
clouds lies at a distance of $\sim$ 50pc at a galactic latitude of $\it b$ $\sim$ 30$^{\circ}$ towards HD 145570 and it is
an extension of the more distant dense wall to the LC in that direction.
The three remaining  neutral clouds appear as a
string of gaseous `blobs'  in the direction of
($\it l$ $\sim$ 165$^{\circ}$, $\it b$ $\sim$ -75$^{\circ}$) at
distances of 60pc (towards HD 9672), 110pc and 180pc.
The cloud
of cold neutral gas at 110pc can be associated with the translucent molecular cloud
G192-67 (Grant $\&$ Burrows \cite{grant99}). This string of 3 nearby clouds would be
an ideal candidate for future soft X-ray shadow observations that potentially could resolve
 the outstanding issue of the degree of pervasiveness of the purported million degree gas
 that may be present within the Local Cavity (Welsh and Shelton \cite{welsh09}). 
 Finally, Figure 13 also shows a thin and elongated cloud of neutral gas at
 a distance of $\sim$ 200pc
 in the galactic direction of ($\it l$ $\sim$ 335$^{\circ}$, $\it b$ $\sim$ -30$^{\circ}$),
 whose existence is revealed for the first time.

\subsection{NaI absorption viewed in the rotational plane}
Figure 14 is perhaps the most changed of all three neutral density maps when compared to the previous
presentations of Paper 1, due
to the small number of target stars available along this plane in the previous work.
The low neutral density rarefied LC region extends to $\sim$ 60pc in most directions, and 
the surrounding fragmented wall of  much denser neutral gas is generally not encountered until a distance of $\sim$ 100pc.
As also revealed in the meridian projection of Figure 13, there is no continuous neutral boundary to the LC at
high latitudes in both galactic hemispheres. In the rotational projection we
see that the Local Chimney has only one major narrow opening
towards the north galactic pole (tilted at $\it b$ $\sim$+75$^{\circ}$), whereas the opening into
the galactic halo in the southern hemisphere is of far larger dimensions. Figure 14 also shows
narrow extensions of the rarefied inner LC to distances as far as $\sim$ 200pc in several directions.
Their finger-like shape means their distances are not well constrained, but it is interesting
to try and relate them to known distant features.
The most prominent of these cavity extensions are seen towards the southern galactic pole
and also towards $\it l$ $\sim$ 250$^{\circ}$ close to the galactic plane. The latter feature
is the pathway
to the GSH 238+00+09 supershell discussed previously in Section 4.1.

Figure 14 suggests that there are no large accumulations of neutral gas contained within the
confines of the rarefied LC in this projection. The nearest  edge of a NaI cloud is at a distance of $\sim$ 40pc in
the direction of $\it l$ $\sim$ 90$^{\circ}$ near to the galactic plane (towards the stars HD 184006
and HD 186882). This cloud appears to be an extension of the denser neutral wall to the LC
in this galactic direction.
We further note the presence of two denser NaI clouds at distances of $\sim$ 85pc and 110pc at galactic
latitude $\it b$ $\sim$75$^{\circ}$. Both clouds are also probably  fragments of the surrounding disk gas that
may have been disrupted by the expansion of the original Local Bubble into the overlying less
dense galactic halo. 

One particularly interesting feature shown in this map is the
circular region of low neutral gas density at
a distance of $\sim$140pc that lies 25$^{\circ}$ above the galactic plane in the general direction
of ($\it l$ $\sim$ 90$^{\circ}$). The low level of absorption along this sight-line was
first reported by Lilienthal et al. (\cite{lil91}) and we now confirm the presence of a 
large interstellar cavity with similar dimensions to that of the Loop I superbubble feature.

For completeness we note that Meyer et al. (\cite{meyer06}) have reported the detection of a very
cold (T $\sim$ 20K) neutral NaI cloud located at
a distance of $\sim$ 45 - 55pc in
the direction ($\it l$ $\sim$ 230$^{\circ}$, $\it b$ $\sim$ +45$^{\circ}$). Unfortunately
the galactic position of this cloud is such that it does not appear in any of the 3 galactic
projections we have currently presented.
However, this is an important detection since the cloud's presence within
the Local Cavity has been posed as a challenge
to current theories concerning the survival of cold gas clouds that may be immersed in a
purportedly hot (million K) surrounding
plasma.

\section{CaII absorption within 300pc}
In the following sub-sections we now present three maps of the spatial distribution of
CaII absorption within 300pc of the Sun, formed from a 3-D CaII  volume density cube derived in a similar manner
to that described for the NaI data. The new
maps, derived with a correlation length of 20pc, are shown in Figures 15 - 17 and  have the same plotting symbols as those presented for NaI, 
except that the CaII volume density 
iso-contours (yellow, green, turquoise and blue)
correspond to values of log n$_{CaII}$ = -9.9, -9.5, -8.9 and -8.2 cm$^{-3}$ .

\subsection{CaII: A view from above the galactic plane}
As was the case for the NaI data, this map has the largest number of sight-line measurements of CaII and
thus is the most accurate of the 3 different galactic projections all presented here for the first time.
The overall appearance of Figure 15 shares a gross similarity
with the equivalent plot for NaI absorption in that it reveals a central region of mainly low density CaII gas that is surrounded by
a highly fragmented wall of denser partially ionized gas clouds. However,
on closer examination of Figure 15  we
see that (in common with the NaI data) the LC is not one large volume of rarefied low neutral density gas, but in fact
it can be better described as a collection of several smaller
low density cavities that are each surrounded by thin gas
filaments of slightly higher density neutral and/or partially ionized material. 
Although some of these filamentary structures may be artifacts caused
by uncertainties in the inversion mapping method
when only a few target measurements are present along an given sight-line, this is not the case for all of the diffuse gas filaments that appear to be present in Figure 15. Clearly, 
although more measurements will be required to confirm the reality of a collection of cell-like cavity structures
that may be present within
the LC, such a scenario would have some intriguing astrophysical ramifications.
For example, how could such small interstellar
cell-like cavities form within the Local Cavity? One possibility is that each of the rarefied sub-cavities could be
a small ionized stellar HII region (with a neutral/partially ionized
filamentary gas envelope) formed by the stellar wind action of nearby early B-type stars (and to a lesser extent
by local hot white dwarfs). We note that
there are at least 11 stars of spectral type earlier than B3V located within 100pc of the Sun whose combined stellar
UV flux may be responsible for photo-ionization of the surrounding low density medium.
This scenario would provide a natural explanation for the observed ionization state
of the local interstellar medium as measured over many local sight-lines by Lehner et al. (\cite{lehn03}). 
We note that previous estimates of the ionization sources of the local ISM have assumed
the presence of a hot million degree gas in addition to  photoionization from the early-type star $\epsilon$ CMa
in order to explain the physical state of the local interstellar cloud (Slavin $\&$ Frisch \cite{slav08}).
Unfortunately a detailed calculation and comparison of the local ionization field is beyond the scope
of this present work, and we urge theorists in this field to test
whether the picture of a Local Cavity with no million degree gas (Welsh $\&$ Shelton \cite{welsh09}),
but instead containing a collection of small ionized HII
region cavities can reproduce the UV absorption line strengths and velocity structure widely observed along
local sight-lines by Lehner et al. (\cite{lehn03}) and Welsh $\&$ Lallement (\cite{welsh05}).

In contrast with the equivalent absorption map of NaI, the central region of the LC has 
at least six condensations of appreciable CaII absorption (i.e. clouds of
of partially ionized gas) lying within 80pc of the Sun.
The nearest region of CaII absorption is the complex of local interstellar
clouds that extend to $\sim$ 15pc mostly in directions contained within galactic quadrants 1 and 4.
Redfield $\&$ Linsky (\cite{redfield08}), using both ground-based and UV absorption observations recorded
towards nearby stars, have presented more detailed maps of each of the several clouds that
contribute to this local cloud complex and argue that the Sun
is located in a transition zone between two of these clouds.
Our results are in quantitative agreement with their results,
with the majority of this local CaII absorption being attributed to the presence of
the local interstellar cloud, the G-cloud 
and the Blue cloud (Lallement et al. \cite{lall95}, Redfield $\&$ Linsky \cite{redfield08}).
The second nearest region of appreciable CaII absorption is located at a distance of $\sim$50pc in
the general direction of ($\it l$ $\sim$80$^{\circ}$) towards the
stars HD 186882 and HD 193369. This feature is also seen in the corresponding NaI
absorption map as a very low neutral density finger-like gas cloud that looks like an extension of a larger and
denser interstellar cloud located
at $\sim$ 100pc in the same galactic direction. The third nearest region of CaII absorption is located
at $\sim$ 55pc in the direction of ($\it l$ $\sim$325$^{\circ}$) towards the stars HD 134481, HD 134482
and HD 141194. It has no apparent NaI counterpart. The fourth nearest region of CaII
absorption is located at a distance of $\sim$ 60pc in the direction of ($\it l$ $\sim$ 20$^{\circ}$) and this
may either be an extension of the local interstellar cloud complex in this direction or a protrusion from
the surrounding wall to the LC. Finally, there are two regions of appreciable CaII absorption
located at about 70pc towards
$\it l$ $\sim$ 265$^{\circ}$ and
$\it l$ $\sim$ 170$^{\circ}$. While the former cloud has a slightly more distant NaI counterpart, 
none is clearly seen for the second regions.
This gas may be strongly photo-ionized, as is the case for more distant gas seen in this direction (see
next paragraph).

Figure 15 shows that 
the dense CaII (and NaI) cloud structures that surround the LC are accompanied by many 
narrow pathways (tunnels) of lower CaII gas density that extend
into surrounding interstellar regions. Many of the dense regions of NaI absorption that surround the LC close to the
galactic plane (as shown in Figure 11)
are also seen as high density CaII-bearing clouds at the same physical locations. In addition, most of the
interstellar tunnels revealed by the paucity of NaI absorption in Figure 12 (and discussed in Section 4.1) are also
observed as low density regions in the CaII absorption map of Figure 15. The major exception to this is
the low neutral density CMa tunnel  ($\it l$ = 226$^{\circ}$) that is so prominent in the NaI
maps. Figure 15 reveals this direction to
possess a high level of  CaII gas density, which is to be expected if this tunnel of interstellar gas is photo-ionized
by the two early-type stars of $\epsilon$ and $\beta$ CMa (Vallerga \cite{vall98}). 
We also note that the
narrow interstellar tunnel of low neutral gas density leading into the
more distant cavity at ($\it l$ $\sim$ 345$^{\circ}$) is significantly more
obscured in the CaII maps. In contrast, the region lying beyond 150pc in
this direction is devoid of both neutral and partially ionized gas, as one
might expect if the cavity (which we associated with the Loop I superbubble)
is hot and highly ionized.

The main conclusion from Figure 15 is that there are a significant number of partially ionized diffuse gas clouds
that are present within the LC. Their ubiquity, current state of ionization and gas temperature have yet to be fully accounted
by any model of the local ISM (Slavin $\&$ Frisch \cite{slav08}, Vallerga \cite{vall98}, Bruhweiler $\&$ Cheng \cite{bruh88}),
and thus we urge a re-assement of some of these models in the light of the recent findings of the lack of a local million degree
gas and the possible existence of cell-like cavity structures within the LC.

\subsection{CaII absorption viewed in the meridian plane}
In Figure 16 we show the spatial distribution of CaII absorption within 300pc
as projected in the galactic meridian plane. The central region of low CaII 
absorption density essentially mimics the
Local Chimney feature shown in Figure 13 for NaI absorption, but the corresponding
CaII Chimney has a narrower diameter of $\sim$ 100pc and reveals several
small cell-like structures within the rarefied central region.
The LC (when viewed in
CaII)  is also seen to be open-ended towards both high
and low galactic latitudes, with a larger (fragmented) opening towards the southern hemisphere, in concert
with the corresponding map of NaI neutral gas absorption. The central low density volume of the LC is surrounded
in the galactic plane by dense partially ionized CaII gas clouds mostly located within the galactic disk. Many of these clouds
are also seen in the equivalent
NaI absorption map of Figure 13. However, we note that the
narrowing of the Chimney in Figure 16 is mainly
due to the presence of dense CaII clouds located in galactic quadrants 2 and 3 whose
full extent are not traced by NaI absorption. For example, the string of 3 gaseous NaI `blobs'  seen in the direction of
($\it l$ $\sim$ 165$^{\circ}$, $\it b$ $\sim$ -75$^{\circ}$) in Figure 12 have a CaII
counterpart that has the appearance of a large hook-like cloud that comes as close as
$\sim$ 40pc to the Sun. 

There are 3 major condensations of
partially ionized CaII gas lying within
100pc of the Sun in this projection. The nearest of these is the local
cloud complex (d $<$ 15pc) that partially surrounds the Sun and has been discussed in the previous
section.  We note that, as suggested by the galactic plane projection map, this cloud appears to be
physically linked to the wall of denser gas that surrounds the LC region in the direction of the galactic center.
The second closest  cloud of high CaII gas density lies at a distance of $\sim$70pc 
in the direction of the north galactic pole. This cloud has no apparent NaI counterpart.
The third closest dense cloud of CaII is located at a distance of $\sim$ 85pc with a galactic
latitude, $\it b$ $\sim$ -50$^{\circ}$, and again it has no apparent counterpart in the equivalent
NaI absorption map.  

In the direction of the south galactic pole we find a region of high CaII absorption
at a distance of $\sim$ 130pc, whose  position is coincident with that of a
finger-like extension of low density neutral gas seen in Figure 13. On the other hand,  the
large circular CaII cloud in the general direction of $\it l$ $\sim$ 180$^{\circ}$ and distance $\sim$ 150pc with
a galactic latitude of $\it b$ $\sim$ +45$^{\circ}$
has no NaI counterpart.  Due to the lack
of suitable targets with distances $>$ 200pc at high galactic latitudes, we are unable to trace the
distribution of partially ionized gas further into the lower galactic halo. Clearly, additional observations of
fainter and more distant early-type stars will be required in order to accomplish this task.

\subsection{CaII absorption viewed in the rotational plane}
Figure 17 shows the spatial distribution of CaII absorption within 300pc in the galactic rotational plane
projection. We observe a large circular volume of very low CaII absorption density that surrounds the
Sun to a distance of $\sim$ 80pc in most directions. This low density cavity has an 80pc wide
opening into the overlying galactic halo at high positive latitudes, but there is only a narrow
tunnel of low CaII gas density that leads into the inner halo towards the southern
galactic pole.
Within the central low density region there are only two clouds
of high density CaII absorption, both of which seem to be physically part of a bar of partially
ionized gas that includes the local interstellar cloud (within which the Sun is partially immersed). 
One of the high density regions of partially ionized gas is located at $\sim$ 50pc in the galactic plane
in the direction of $\it l$ $\sim$ 90$^{\circ}$, and can be identified with the spatially
coincident gas cloud detected in NaI
and discussed in Section 4.3. The other local high density CaII cloud is in the general direction of $\it l$ $\sim$ 270$^{\circ}$
at a distance of $\sim$ 35pc, located just below the Sun in Figure 17.
This cloud complex can be identified
with absorption due to both the interstellar G-cloud and Auriga cloud 
as listed by Redfield $\&$ Linsky (\cite{redfield08}). We also note that the bar of CaII gas within
the LC seems to be
linked to the surrounding boundary  wall (at least in the direction of $\it l$ $\sim$ 90$^{\circ}$). 
Its physical shape and dimensions are suggestive of a cell-like structure, in agreement
with the local web of low-density interstellar sub-cavities discussed in Section 5.1.

The fragmented wall of high CaII gas density that 
mostly surrounds the central LC region is
generally matched with spatially coincident regions of high NaI neutral gas density. However, this
one-to-one matching is not always observed. For example,
the tunnel of $\it low$ neutral density gas that opens into the galactic halo towards the north
galactic pole shown in Figure 14 is seen to be spatially coincident with a
$\it high$ density region of CaII absorption. Similarly, the large high density CaII
cloud at 110pc seen towards the south galactic pole is matched by a similarly
large region of very low NaI gas density.
This pattern of NaI and CaII
absorption behavior is reflected in the derived
NaI/CaII column density ratio, whose major variations within 300pc are discussed in the following section.

\section{Variation of the NaI/CaII ratio}
The column density ratio of N(NaI)/N(CaII) is a well-known diagnostic of the physical
conditions in the diffuse interstellar gas, since Ca is more sensitive than Na to the
balance between adsorption on, and desorption from, interstellar grains (Barlow \cite{barl78}).
In cold (T $\sim$ 30K) and dense gas clouds, in which most of the gas-phase Ca is
depleted onto grains, the NaI/CaII ratio is $>$ 100. However,  in the warmer (T $\sim$ 1000K)
and lower density ISM where much of the Ca remains in the gas phase, ratios of $<$ 1.0
are commonly found (Hobbs \cite{hobbs75}, Centurion $\&$ Vladilo \cite{cent91}, Bertin at al. \cite{bertin93}).
For sight-line distances $\la$ 30pc the NaI/CaII column density ratio for the warm (T $\sim$ 7000K) local 
interstellar gas clouds is
$\sim$ 0.2 (Bertin et al. \cite{bertin93}). Equilibrium equations show that
at 7000K the NaI/CaII ratio is independent of electron density
and as a consequence this  ratio becomes a tracer of the level of calcium elemental depletion
for the local gas.

Beyond the LC and in the general ISM the NaI/CaII ratio is thought to be a function of gas cloud velocity, such
that at velocities $\ga$ 30 km s$^{-1}$ interstellar dust grains may be destroyed by shocks and the Ca is
liberated into the gas phase (Routly $\&$ Spitzer \cite{rout52}, Siluk $\&$ Silk \cite{siluk74}). Thus, as
has been argued by Crawford et al. (\cite{craw02}), a low value of the NaI/CaII ratio could therefore be due to
either the presence of warm and partially ionized gas and/or the presence of interstellar shocks. Since we do not
expect, nor have observed, interstellar gas with velocities $\ga$ 30 km s$^{-1}$ in the local ISM then for distances $\la$
100pc any variation in the observed NaI/CaII ratio is most probably due to temperature/ionization effects. 
From our sample of stars that possess measured values
of both NaI and CaII column densities we see that there is a range in NaI/CaII ratio
values of 0.03 to 100, with the vast majority of sight-lines having ratio values $<$ 5.0. 
Since the column density values listed in Table 1 are total integrated values measured over
the entire length of each sight-line, the derived NaI/CaII ratio value is a sight-line average value. Thus, variations
in the ratio for low column density cloud components that may be present along a particular sight-line will be masked by
the column density contribution from any high column density (cold and dense) regions. 
One example of this is shown in Figure 2 for 
HD 30112, whose total NaI/CaII column density ratio is 2.1 whereas the V $\sim$ 5 km s$^{-1}$ cloud component
has a ratio of only 1.15. Thus, we defer a detailed study of the spatial variation of the NaI/CaII ratio in the local ISM
to the forthcoming paper by Lallement et al. (\cite{lall10}) that will focus on the cloud component velocity structure of these
sight-lines.

In Figure 18 we show plots of the column density ratio of NaI/CaII as a function of distance to 400pc for
sight-lines with latitudes near the galactic plane ($\it b$ $\pm$20$^{\circ}$) and split by longitude into 4 galactic quadrants.
We see that for distances $<$ 80pc the majority of $\it measured$ values of the NaI/CaII ratio
lie in the 0.1 to 1.0 range, in general agreement with the low value for this ratio found by Bertin et al. (\cite{bertin93})
for stars within the Local Cavity. For more distant sight-lines the NaI/CaII is more variable, but in most
directions the ratio value falls in the range 0.5 to 20. This is in agreement with
with previous work on the NaI/CaII ratio in the general ISM by Welty et al. (\cite{welty96}).

The highest values
of the NaI/CaII ratio near the galactic plane are found in
quadrant 2 in the direction of l $\sim$ 150$^{\circ}$ towards the Taurus dark clouds (i.e. along sight-lines towards 
HD 23180 (55.0), HD 23552 (72.4), HD 24398 (85.1) and HD 25642 (102.3). There appears to be
qualitative agreement in the distribution of the NaI/CaII with distance in all 4 galactic quadrants, with
a smaller range of ratio values (typically 0.1 to 5) present in quadrant 3. This is most probably due
to the presence of significant amounts of partially
ionized CaII gas in the 200pc long low neutral density interstellar tunnel seen towards the
stars $\beta$ and $\epsilon$ CMa which shows an ionization gradient in this galactic
direction (Wolff et al. \cite{wolf99}). 

\section{Conclusion}
We have presented new high spectral resolution absorption observations
of the interstellar NaI D-lines (5890\AA) recorded along
482 sight-lines, together with 807 new measurements of the
CaII K-line (3933\AA). When combined with existing data in the literature, we have compiled a catalog
of absorption measurements towards a total of 1857 early-type stars located within 800pc of the Sun (1678 stars
measured at NaI and 1267 stars measured at CaII).
By measuring both the NaI and CaII interstellar line equivalent widths and then subsequently fitting the normalized absorption
profiles with a line velocity, doppler width and column density, we have been able to characterize the neutral and
partially ionized  properties of the interstellar medium along each sight-line. A plot of the equivalent width of NaI D2-line versus distance reveals a central region of very low neutral gas absorption (W$_{\lambda}$(D2) $<$ 5m\AA) out to a distance
of 80pc in most galactic directions. This is the well-known Local Cavity region, which has a wall of neutral NaI gas
(of W$_{\lambda}$(D2) $>$ 50m\AA) beyond 80pc in most directions.
 In contrast, a similar plot for the equivalent width of CaII shows no sharply increasing absorption at 80pc, but instead we observe a slowly increasing value of CaII  equivalent width with increasing sight-line distance sampled.

Low values for the volume density of NaI (n$_{naI}$ $<$10$^{-9}$ cm$^{-3}$) are generally found
within 50pc of the Sun (i.e. within the Local Cavity), whereas
values in the range 10$^{-8}$  $>$ n$_{NaI}$ $>$ 10$^{-10}$ cm$^{-3}$ are found for sight-lines with distance $>$ 300pc.
Both high and low values of the volume density of CaII (n$_{CaII}$) are found for sight-lines $<$ 30pc, dependent on
whether local gas cloudlets are encountered. For distances $>$ 100pc a value of n$_{CaII}$ $\sim$ 10$^{-9}$ cm$^{-3}$
is typical for most sight-lines, suggesting that the density distribution of CaII gas absorption is fairly uniform
throughout the general ISM. 

Each of the normalized absorption profiles of the NaI and CaII lines were then
subsequently fit with models  of line component velocity, doppler width and column density, with
the total integrated values of column density for each ion along each sight-line being presented in Table 1.
These new data were then used in
conjunction with previous absorption measurements published in the literature to construct
3-D maps of the spatial distribution 
of neutral NaI and partially ionizes CaII gas density. This was achieved from the inversion of
each column-density determination using a method based
on the work of  Vergely et al. (\cite{verg01}).
The NaI maps extend and improve upon the accuracy of similar maps constructed by Lallement et al. (\cite{lall03}),
whereas the CaII maps are the first of their kind to be published.

The general appearance of the maps of the 3-D spatial distribution of NaI gas density are similar to those of CaII
absorption, in that we find a $\sim$ 80pc diameter central region of low absorption (i.e. the Local Cavity) surrounded by
a  wall of high absorption density. In contrast
to the preliminary maps of NaI absorption presented in Paper 1, the surrounding wall of NaI (and CaII)
absorption is now found to be highly fragmented with several new interstellar tunnels of low gas density
emanating from the central LC region. This fragmentation could be linked to the origin of the LC, which
is generally thought to have been formed by a SN explosion over several million years ago.
For most sight-lines many regions of high NaI absorption are 
spatially matched by regions of high CaII absorption. However, there are several important differences
within the details of these maps. Firstly, the central region of the Local Cavity contains several clouds of
high CaII gas density, in contrast to the general absence of neutral NaI gas clouds found within the rarefied cavity.
The observed pattern of local CaII absorption appears to be in the form
of a collection of several very low density interstellar cells that are
surrounded by partially ionized CaII filaments. This collection of low density cavities might be caused by the combined
action of stellar winds from nearby early B-type stars that could be responsible for the photoionization of the
local interstellar gas. Secondly, although there are low gas density pathways that lead from the LC into the
overlying galactic halo, the low density neutral NaI gas openings are of far larger dimensions than their low density
CaII equivalents. Thirdly, there are several regions of high NaI absorption that do not have high CaII absorption
counterparts, and vice-versa. This is reflected in the value of the integrated column density ratio of NaI/CaII
for these clouds, which is presumably influenced by the ambient level of ionization in these regions.

Plots of the integrated column density ratio of NaI/CaII as a function of distance for sight-lines
near to the galactic plane have
values that lie in the range 0.1 to 1.0 for sight-lines with distances $<$ 80pc.  However, ratio values
of between 0.5 and 20 are typical for more distant sight-lines. The highest values of the NaI/CaII ratio
are found towards $\it l$ $\sim$ 150$^{\circ}$ in the direction of the Taurus dark clouds, and ratio values
in the narrower range of 0.1 to 5 are found in galactic quadrant 3 due to the presence of the low neutral density
$\beta$ CMa interstellar tunnel.
   
\begin{acknowledgements}
We particularly acknowledge the dedicated
team of engineers, technicians, and research staff at the
Observatoire de Haute Provence, Mt. John Observatory, European Southern
Observatory, South African Astronomical Observatory and the Lick Observatory.
The observations recorded at the European
Southern Observatory were taken as part of the Large Programs 077.C-0575 and
179.c-0197.
This publication makes use of data products from the SIMBAD database,
operated at CDS, Strasbourg, France.We also thank
Christopher Henderson (University of Canterbury, Christchurch, NZ) for his help in
making observations and Dr. Seth Redfield for providing us with his high resolution CaII K-line results
prior to their publication in the literature. We acknowledge the effort of Jonathan
Wheatley who made many of the Figures for this paper and thank the referee for his comments which
greatly improved the Paper. BYW acknowledges funding for this
research through the NSF award AST-0507244.
\end{acknowledgements}

\newpage

\begin{figure}
\center
{\includegraphics[height=18cm]{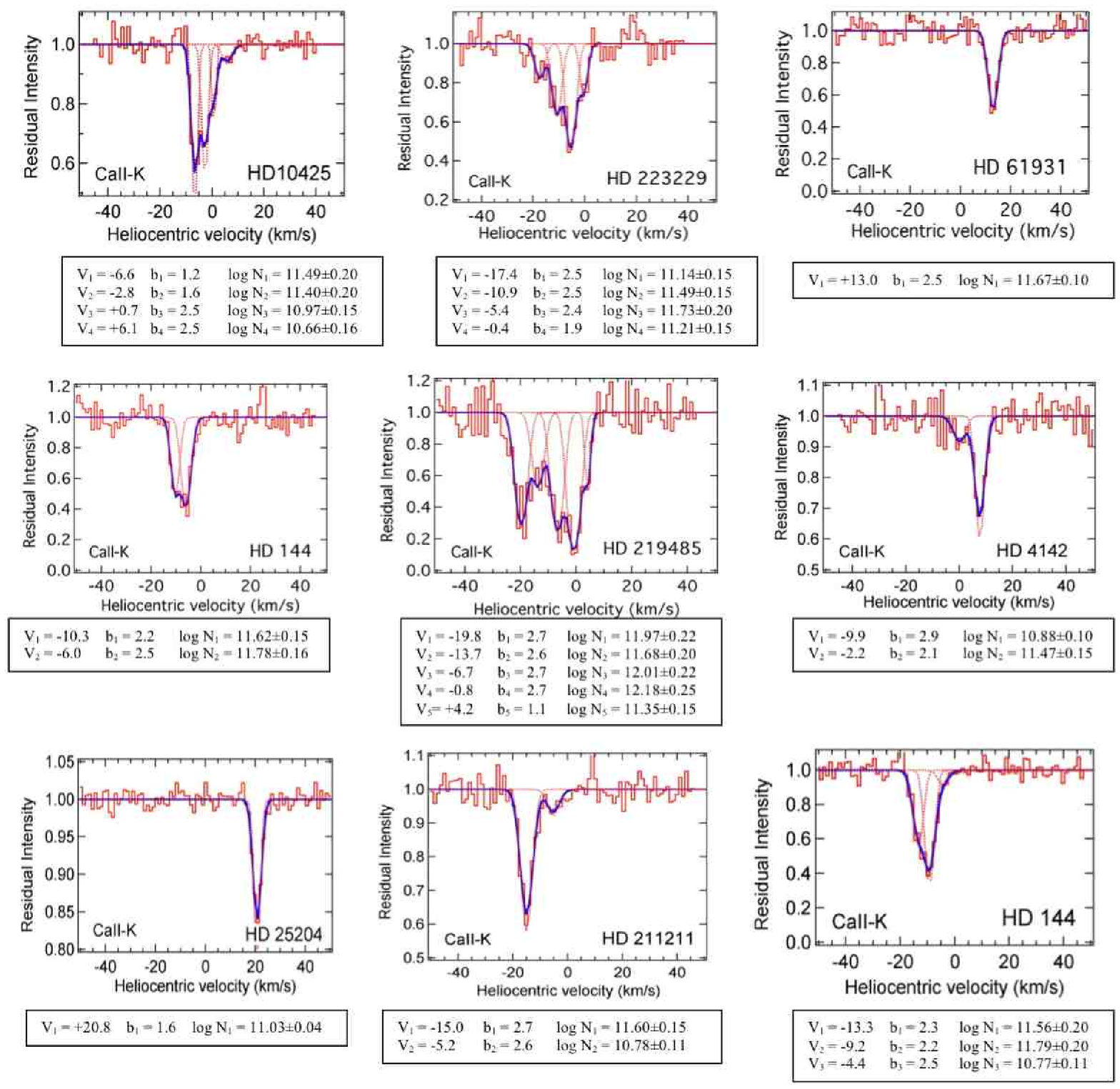}}
\caption{Examples of interstellar CaII-K line absorption profiles together with best-fit models (thick line) recorded with the Aurelia spectrograph on the 1.52m telescope at the Observatoire de Haute Provence. Dotted lines are the model components prior to instrumental convolution. The best-fit component values of velocity (V), doppler-width (b) and column density (N) are listed in the box beneath each profile.}
\label{Figure 1}
\end{figure}
\clearpage

\begin{figure}
\center
{\includegraphics[height=18cm]{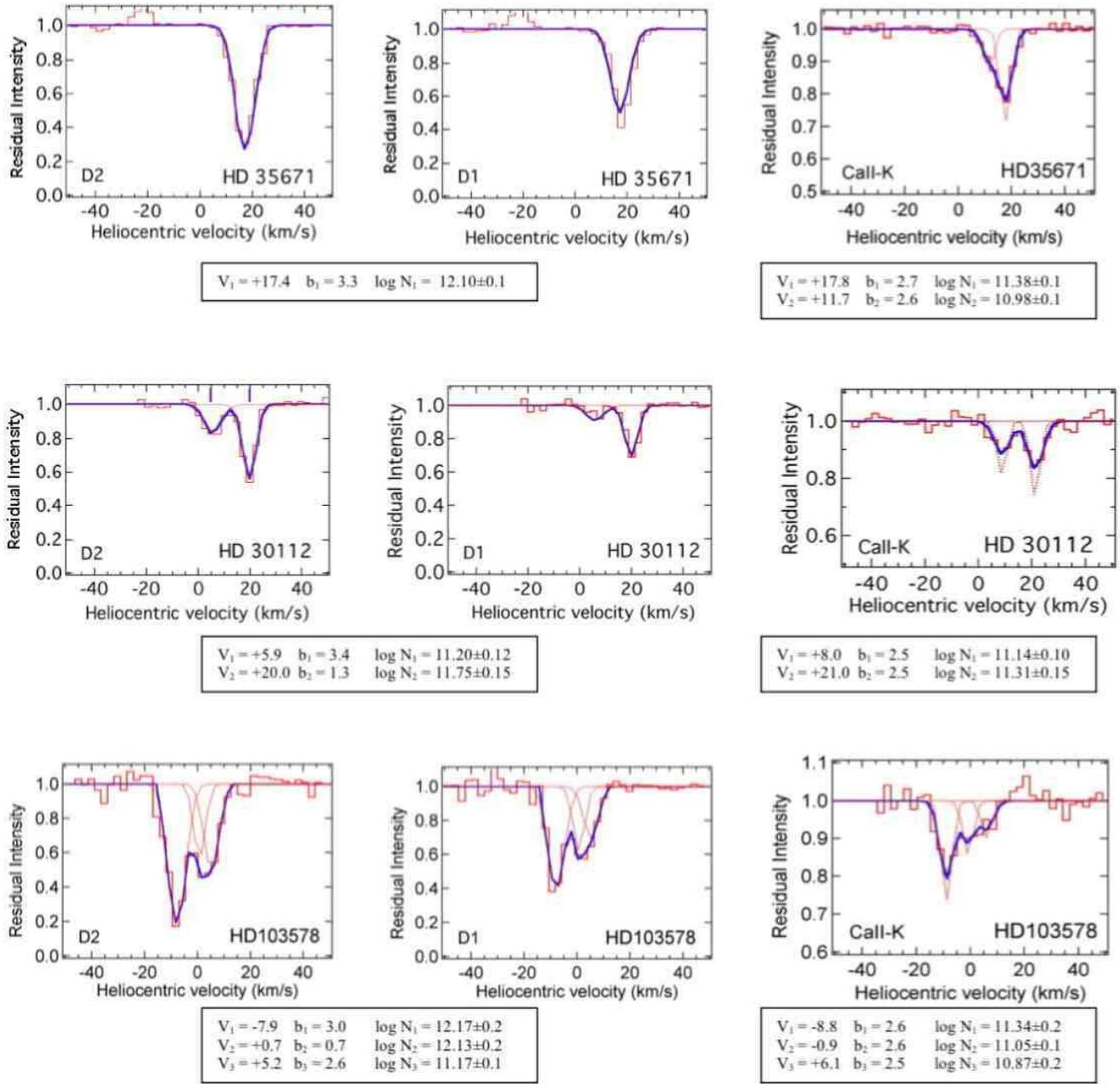}}
\caption{Examples of interstellar NaI D2 and D1 and CaII-K line absorption profiles together with best-fit models (thick line) recorded with the Hamilton spectrograph on the 0.9 m CAT telescope at the Lick Observatory of the University of California. Dotted lines are the model components prior to instrumental convolution. The best-fit component values of velocity (V), doppler-width (b) and column density (N) are listed in the box beneath each profile.}
\label{Figure 2}
\end{figure}
\clearpage

\begin{figure}
\center
{\includegraphics[height=18cm]{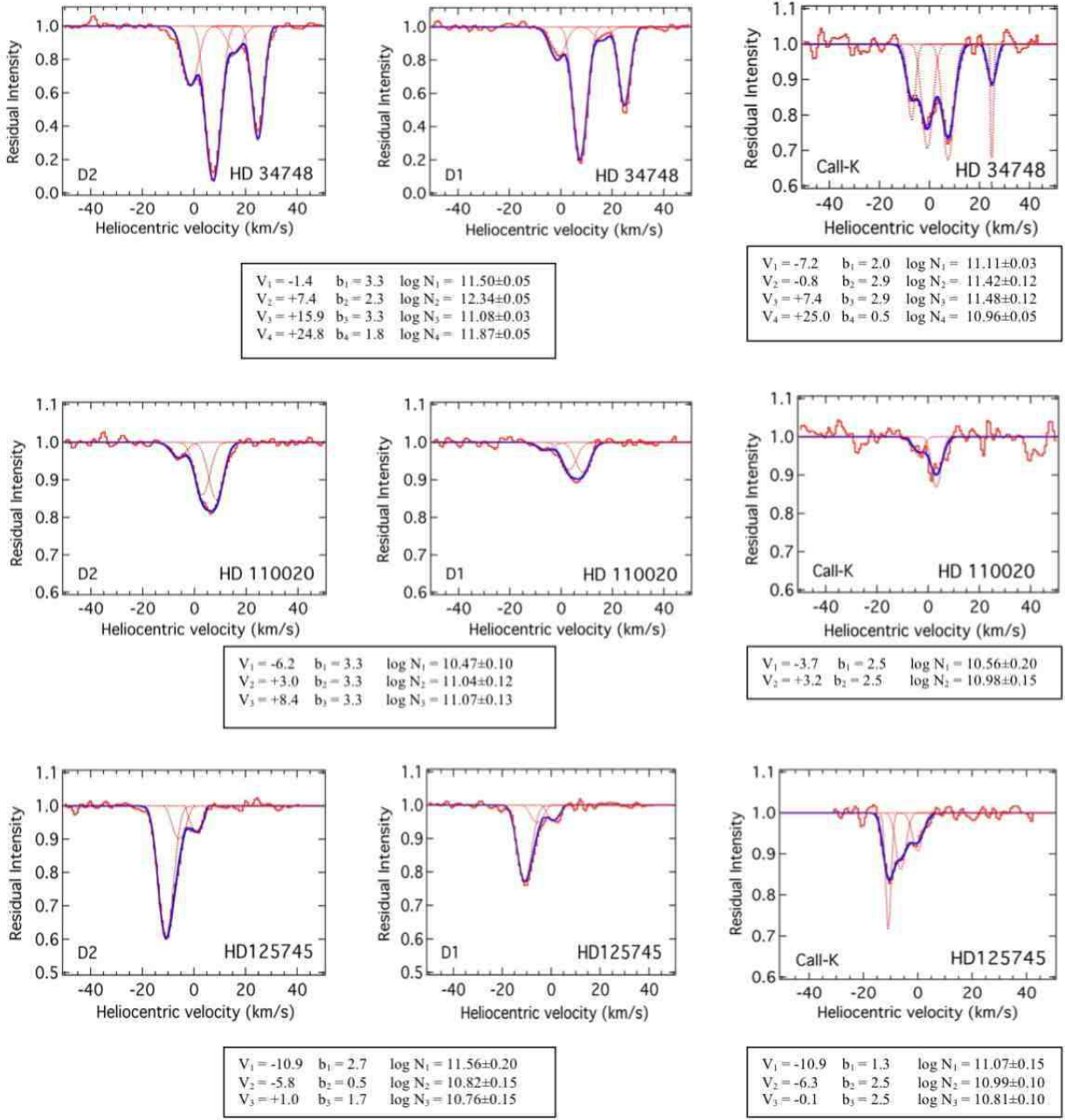}}
\caption{Examples of interstellar NaI D2 and D1 and CaII-K line absorption profiles together with best-fit models (thick line) recorded with the Hercules spectrograph on the 1.0 m telescope at the Mt. John Observatory in New Zealand. Dotted lines are the model components prior to instrumental convolution. The best-fit component values of velocity (V), doppler-width (b) and column density (N) are listed in the box beneath each profile. The short vertical lines above some profiles indicate the position of the absorption component's velocity.}
\label{Figure 3}
\end{figure}
\clearpage

\begin{figure}
\center
{\includegraphics[height=18cm]{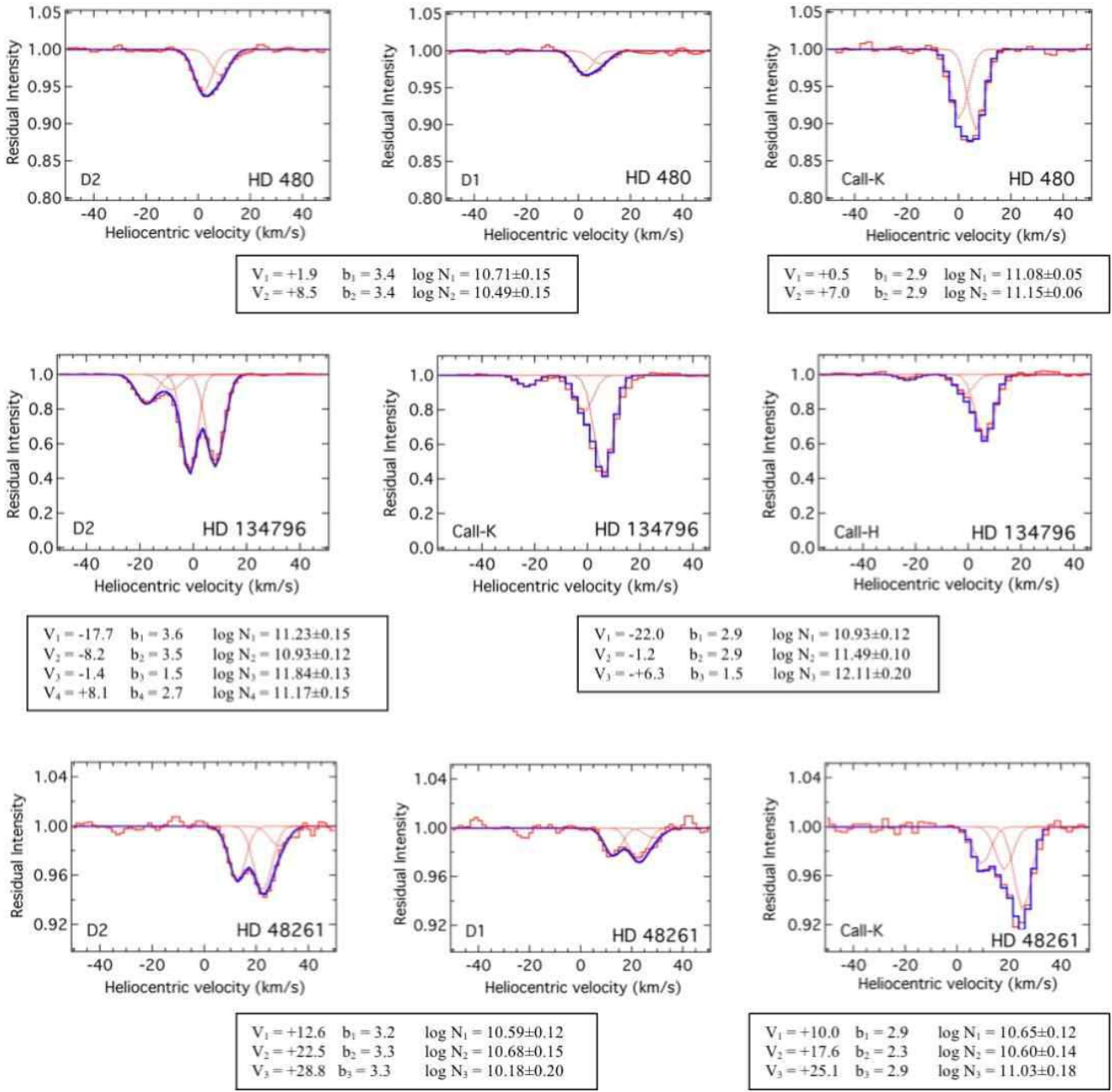}}
\caption{Examples of interstellar NaI D2 and D1 and CaII K and H- line absorption profiles together with best-fit models (thick line) recorded with the FEROS spectrograph on the 2.2 m telescope at the European Southern Observatory in Chile. Dotted lines are the model components prior to instrumental convolution. The best-fit component values of velocity (V), doppler-width (b) and column density (N) are listed in the box beneath each profile.}
\label{Figure 4}
\end{figure}
\clearpage

\begin{figure}
\center
{\includegraphics[height=18cm]{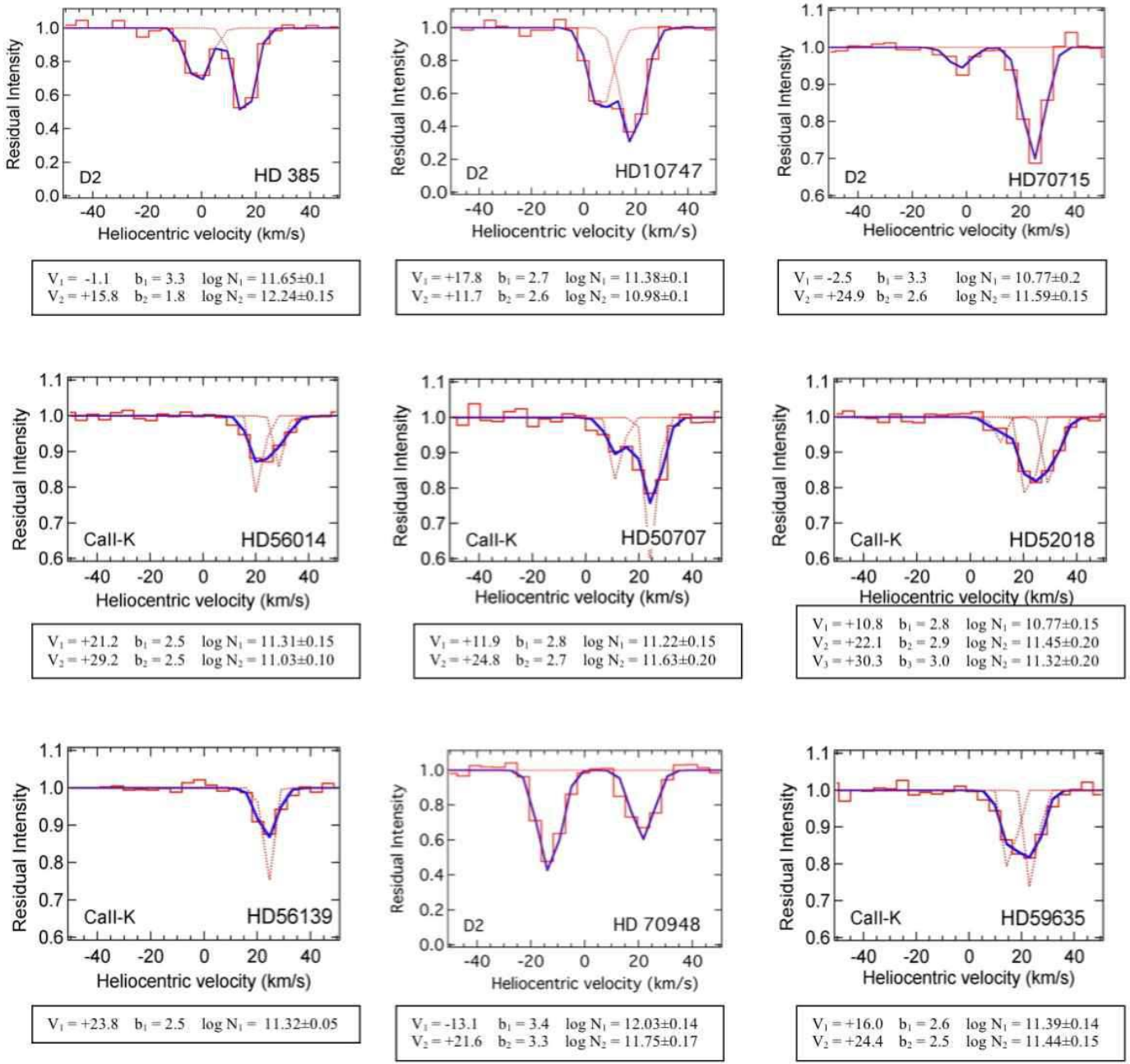}}
\caption{Examples of interstellar NaI D2 and D1 and CaII-K line absorption profiles together with best-fit models (thick line) recorded with the GIRAFFE spectrograph on the 1.9 m telescope at the South African Astronomical Observatory. Dotted lines are the model components prior to instrumental convolution. The best-fit component values of velocity (V), doppler-width (b) and column density (N) are listed in the box beneath each profile.}
\label{Figure 5}
\end{figure}
\clearpage

\begin{figure}
\center
{\includegraphics[height=10cm]{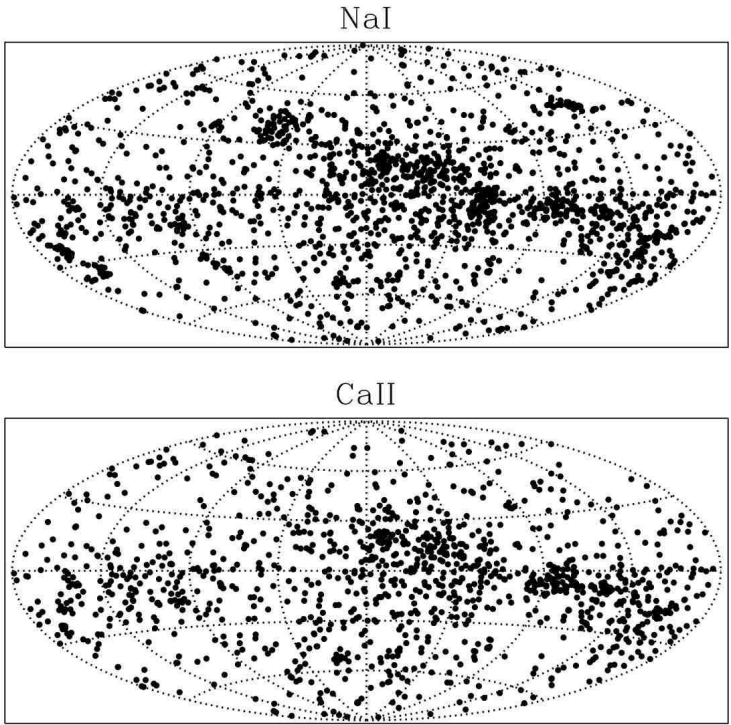}}
\caption{Galactic distribution of  the interstellar sight-lines sampled by (a) NaI absorption, and (b) CaII absorption. Both plots are centered at galactic co-ordinate (0,0).}
\label{Figure 6}
\end{figure}
\clearpage

\begin{figure}
\center
{\includegraphics[height=10cm]{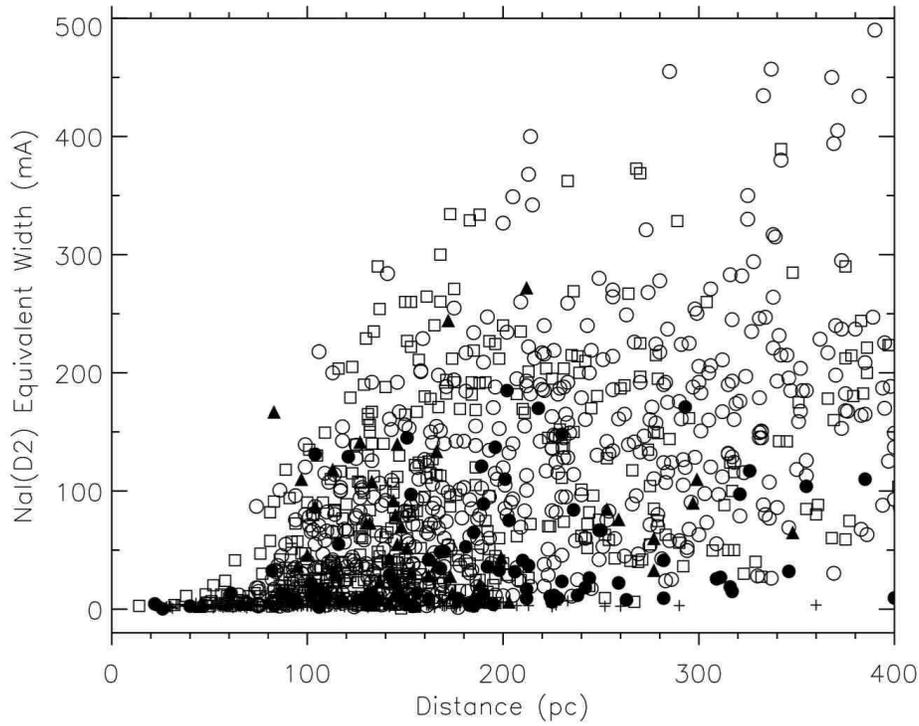}}
\caption{Plot of the equivalent width (m\AA) of the interstellar NaI D2-line for stars with distances $<$ 400pc. Filled triangles are for sight-lines with galactic latitude $\it b$ $>$ +45$^{\circ}$, open squares for sight-lines with $\it b$ = 0 to 45$^{\circ}$, open circles  for sight-lines with $\it b$ = 0 to -45$^{\circ}$ and filled circles for sight-lines with $\it b$ $<$ -45$^{\circ}$. Crosses are upper limit values.  Note the sharp increase in the level of NaI absorption at $\sim$ 80pc, which is due to the neutral wall to the Local Cavity.}
\label{Figure 7}
\end{figure}
\clearpage

\begin{figure}
\center
{\includegraphics[height=10cm]{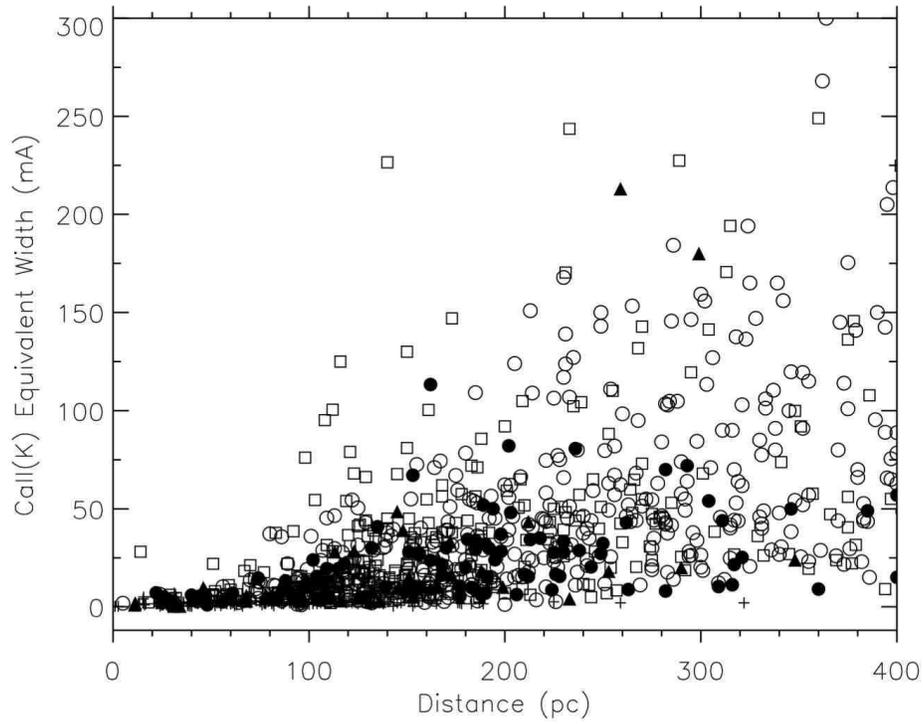}}
\caption{Plot of the equivalent width (m\AA) of the interstellar CaII K-line for stars with distances $<$ 400pc. Symbols are the same as those in Figure 7. }
\label{Figure 8}
\end{figure}
\clearpage

\begin{figure}
\center
{\includegraphics[height=10cm]{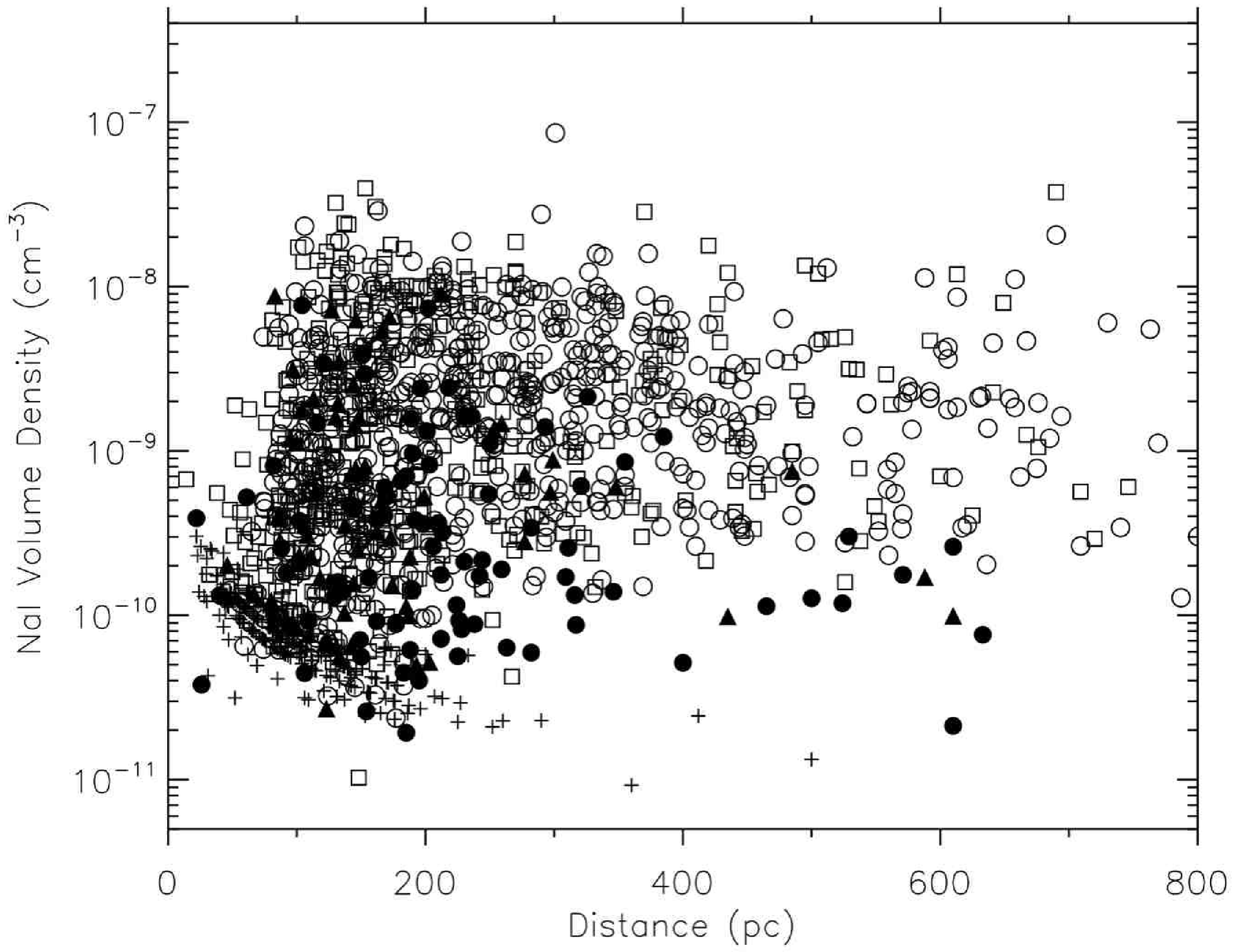}}
\caption{Plot of the volume density of interstellar NaI absorption for stars with distances $<$ 800pc. Symbols are the same as Figure 7. }
\label{Figure 9}
\end{figure}
\clearpage

\begin{figure}
\center
{\includegraphics[height=10cm]{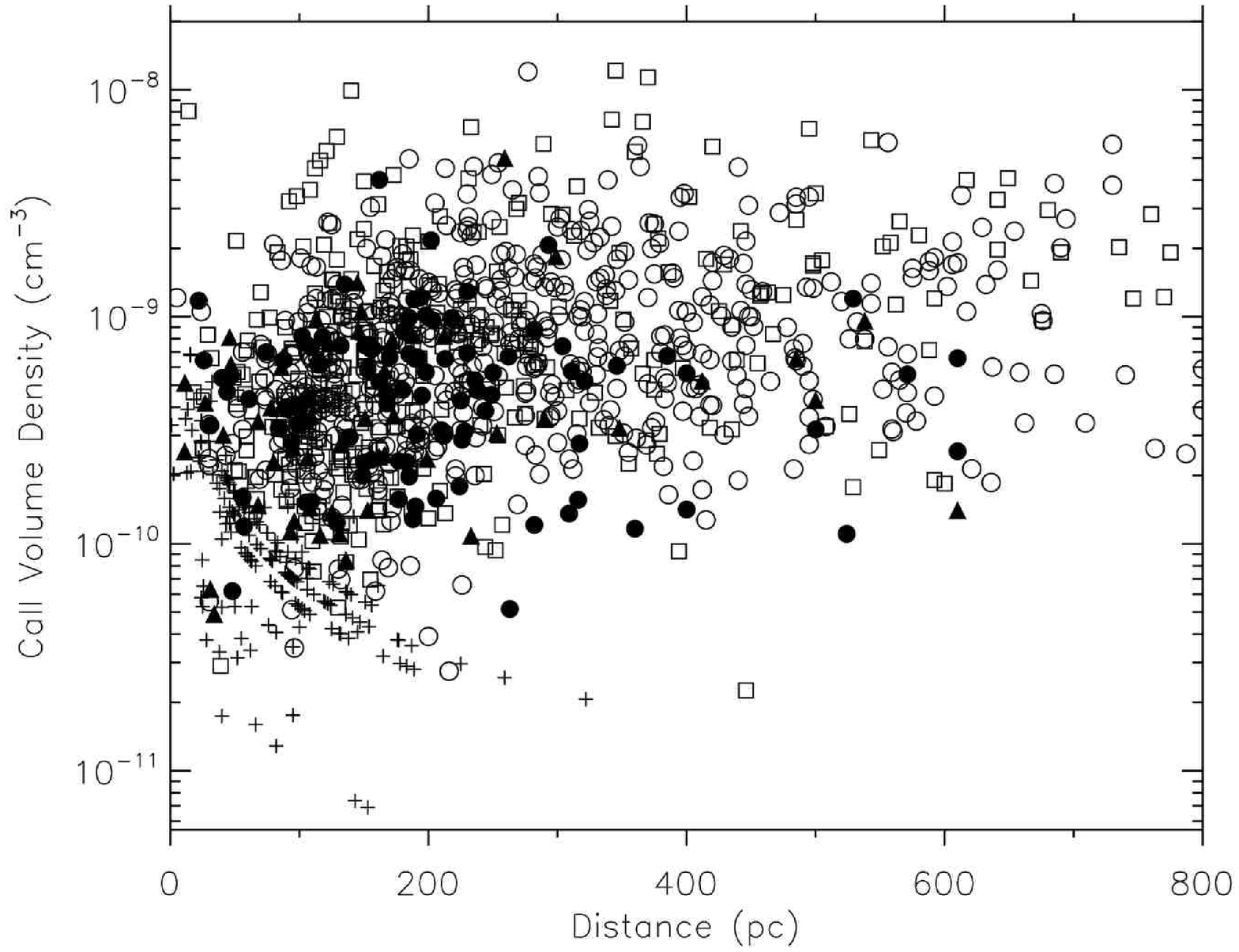}}
\caption{Plot of the volume density of interstellar CaII absorption for stars with distances $<$ 800pc. Symbols are the same as those in Figure 7.}
\label{Figure 10}
\end{figure}
\clearpage

\begin{figure}
\center
{\includegraphics[height=8cm]{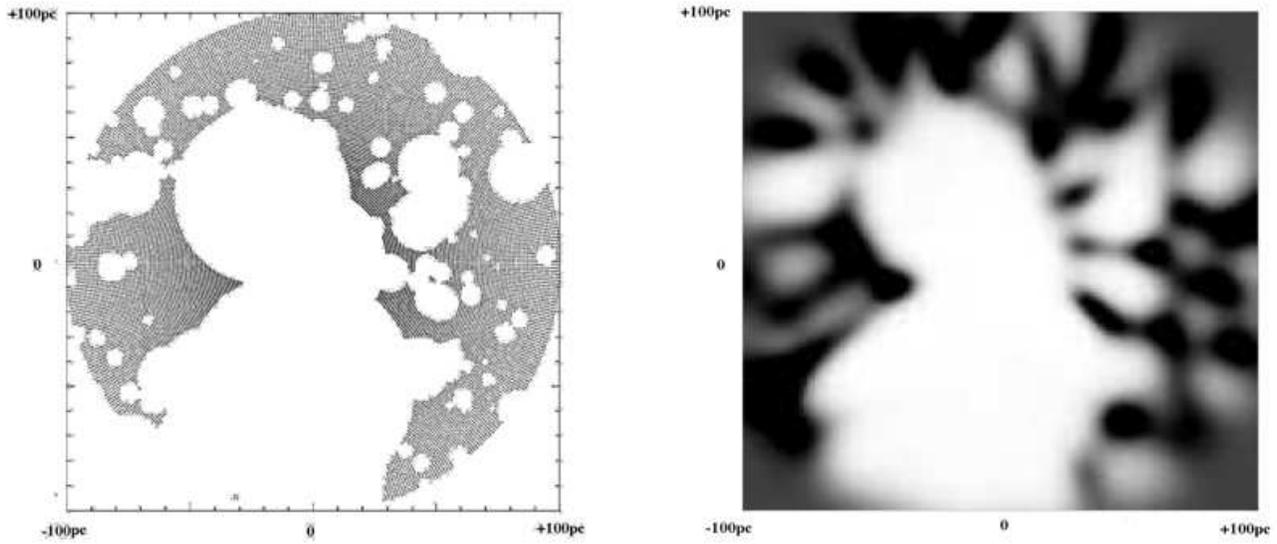}}
\caption{Comparison of  an initial map of the fractal distribution of interstellar reddening in the
galactic plane (left) and the equivalent map of gas column densities (right) derived from the inversion method of Vergely (1998). Most large features are reproduced, whereas small features with sizes $<$ 15pc are smeared by the inversion method.}
\label{Figure 11}
\end{figure}
\clearpage

\begin{figure}
\center
{\includegraphics[height=16cm]{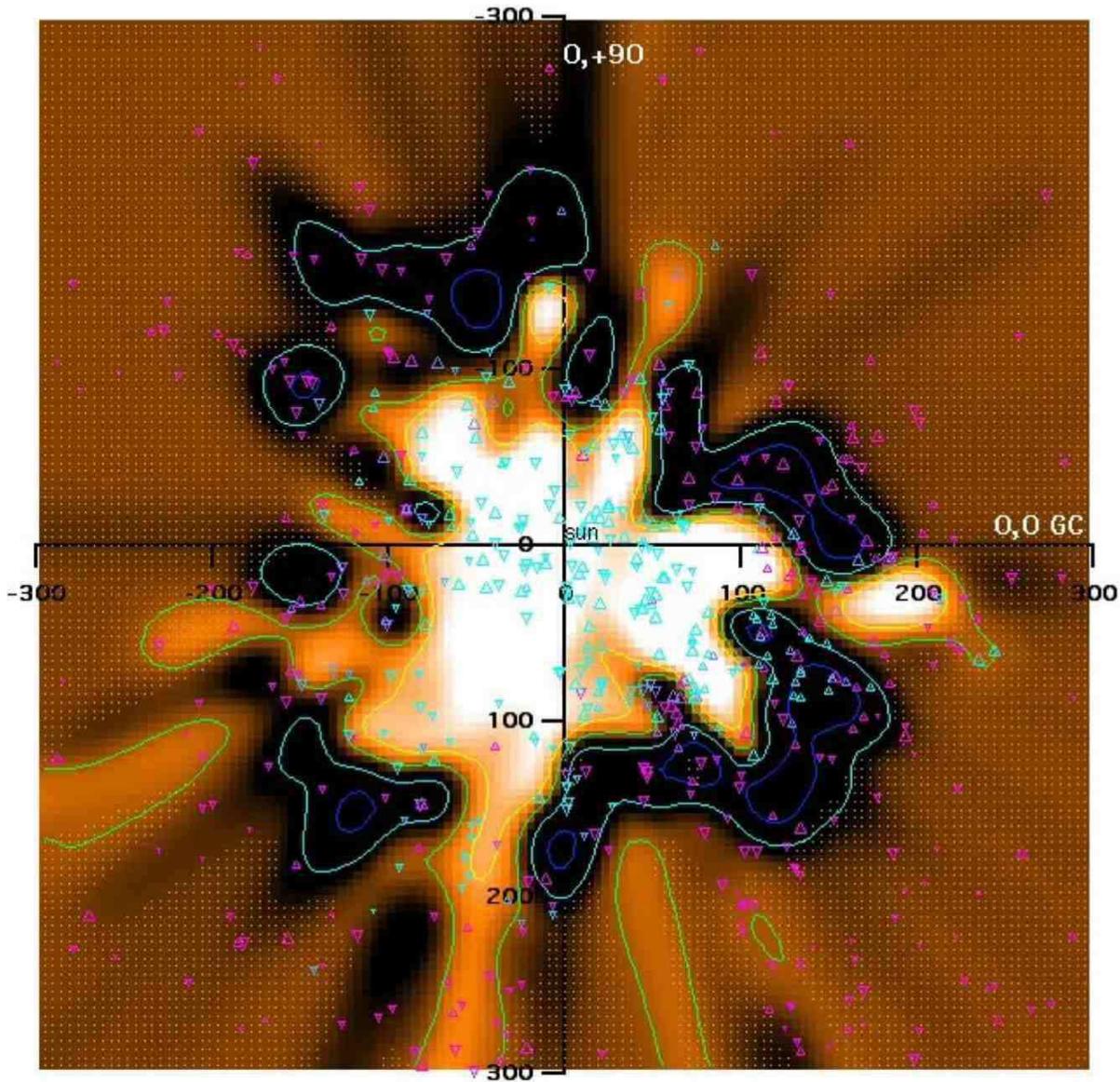}}
\caption{Plot of 3-D spatial distribution of interstellar NaI absorption within 300pc of the Sun as viewed in the galactic plane projection. Triangles represent the sight-line positions of stars used to produce the map, with the size of the triangle being proportional to the derived NaI column density. Stars plotted with vertex upwards are located above the galactic plane, vertex down are below the plane. White to dark shading represents low to high values of the NaI volume density(n$_{NaI}$).  The corresponding iso-contours (yellow, green, turquoise and blue) for log n$_{NaI}$ = -9.5, -9.1, -8.5 and -7.8 cm$^{-3}$ are also shown.  Regions with a matrix of dots  represent areas of uncertain neutral gas density measurement.}
\label{Figure 12}
\end{figure}
\clearpage

\begin{figure}
\center
{\includegraphics[height=16cm]{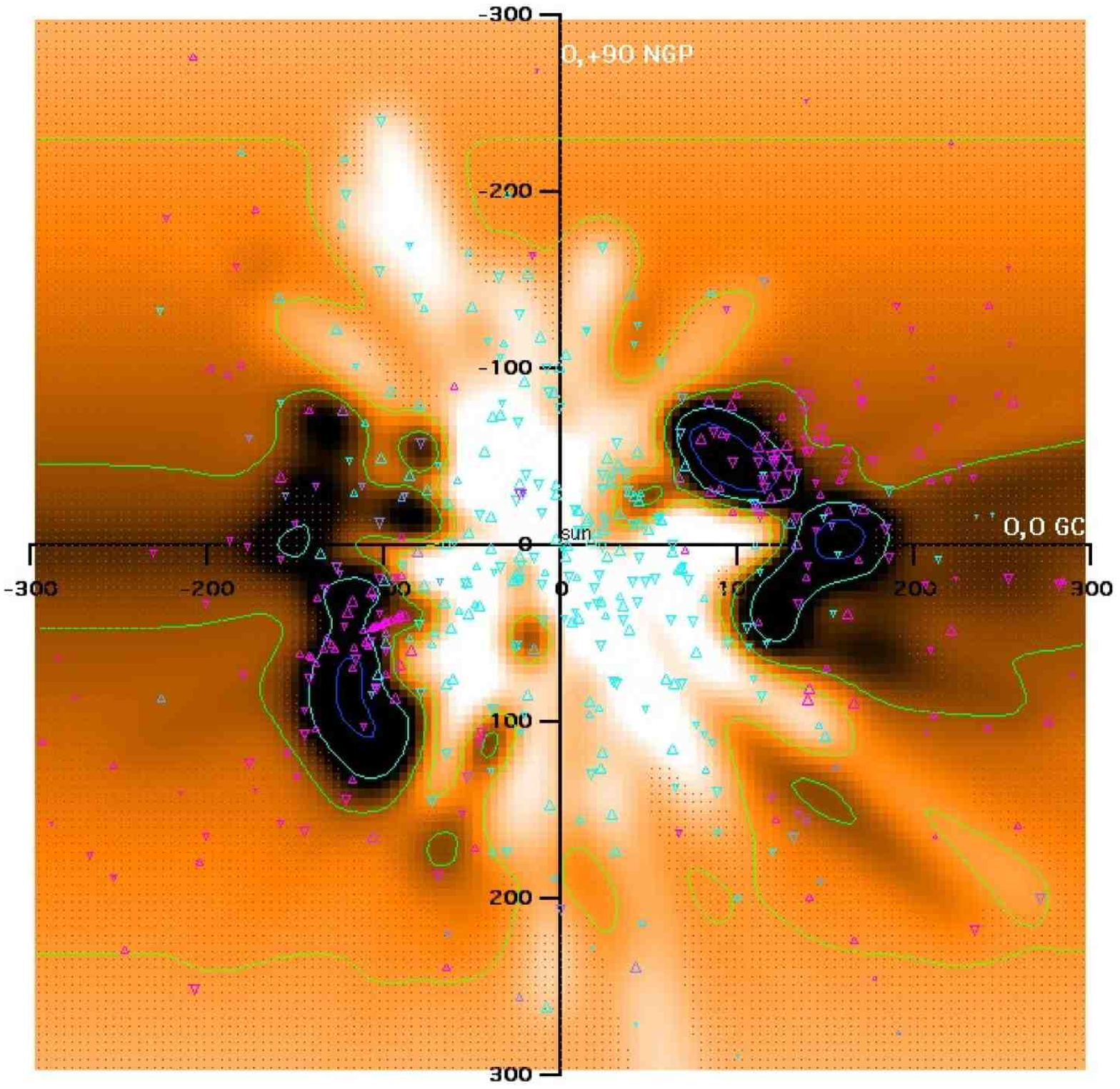}}
\caption{Plot of 3-D spatial distribution of interstellar NaI absorption within 300pc of the Sun as viewed in the meridian plane projection. Plotting symbols are the same as those given for Figure 11.}
\label{Figure 13}
\end{figure}
\clearpage

\begin{figure}
\center
{\includegraphics[height=16cm]{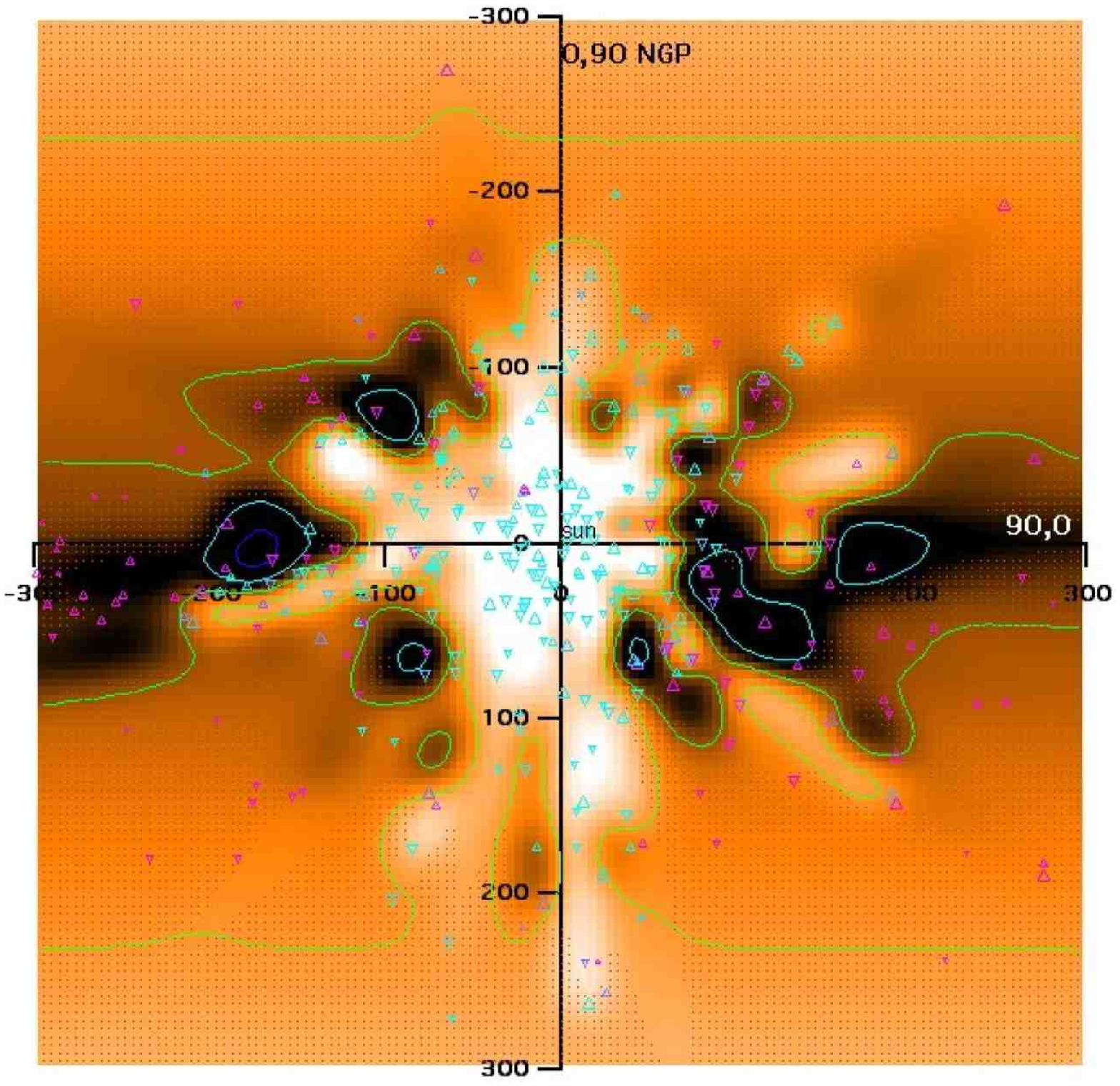}}
\caption{Plot of 3-D spatial distribution of interstellar NaI absorption within 300pc of the Sun as viewed in the rotational plane projection. Plotting symbols are the same as those given for Figure 11.}
\label{Figure 14}
\end{figure}
\clearpage

\begin{figure}
\center
{\includegraphics[height=16cm]{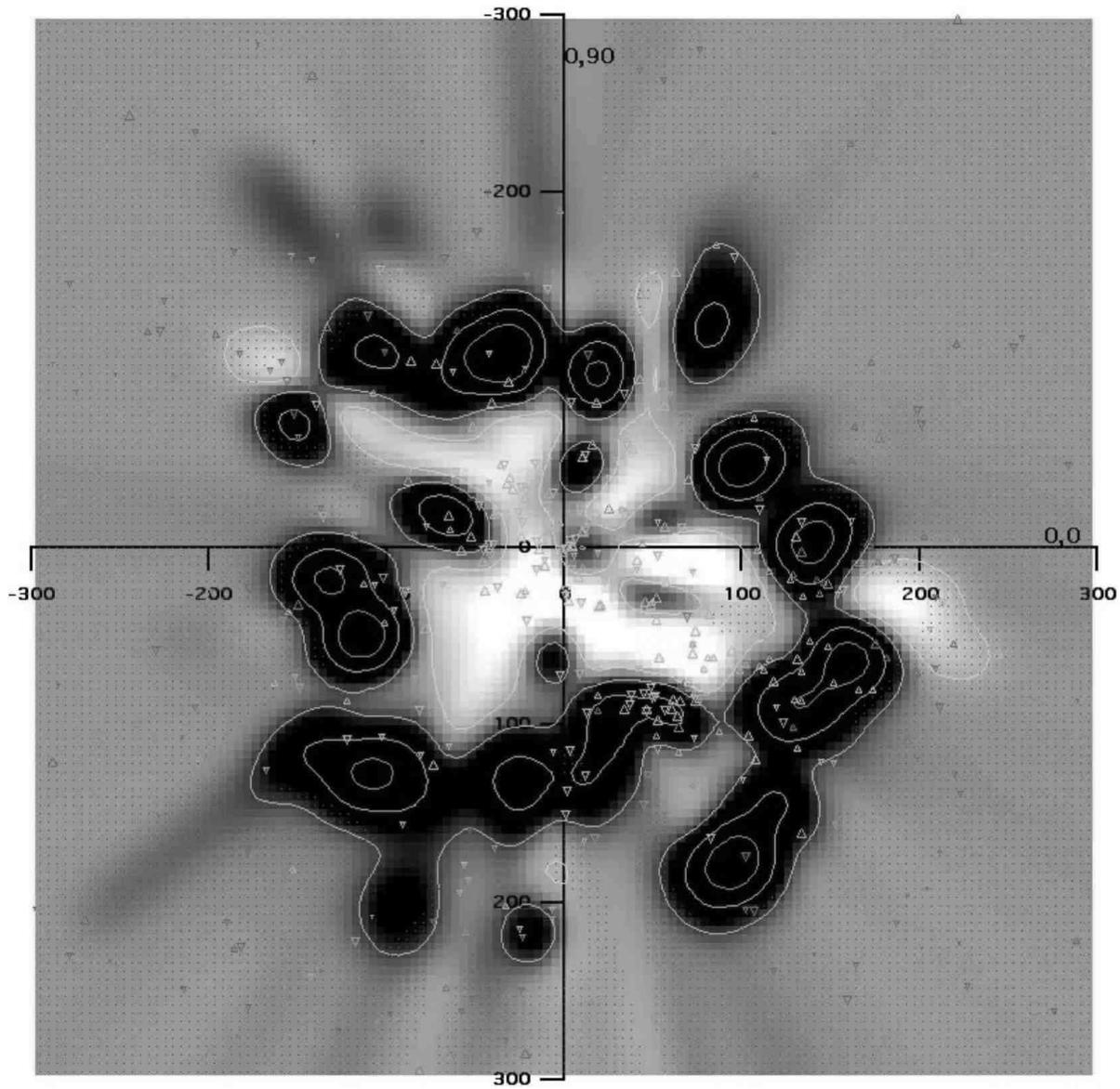}}
\caption{Plot of 3-D spatial distribution of interstellar CaII absorption within 300pc of the Sun as viewed in the galactic plane projection. Plotting symbols are the same as Figure 11, except that the volume density 
iso-contours (yellow, green, turquoise and blue)
correspond to log n$_{CaII}$ = -9.9, -9.5, -8.9 and -8.2 cm$^{-3}$}.
\label{Figure 15}
\end{figure}
\clearpage

\begin{figure}
\center
{\includegraphics[height=16cm]{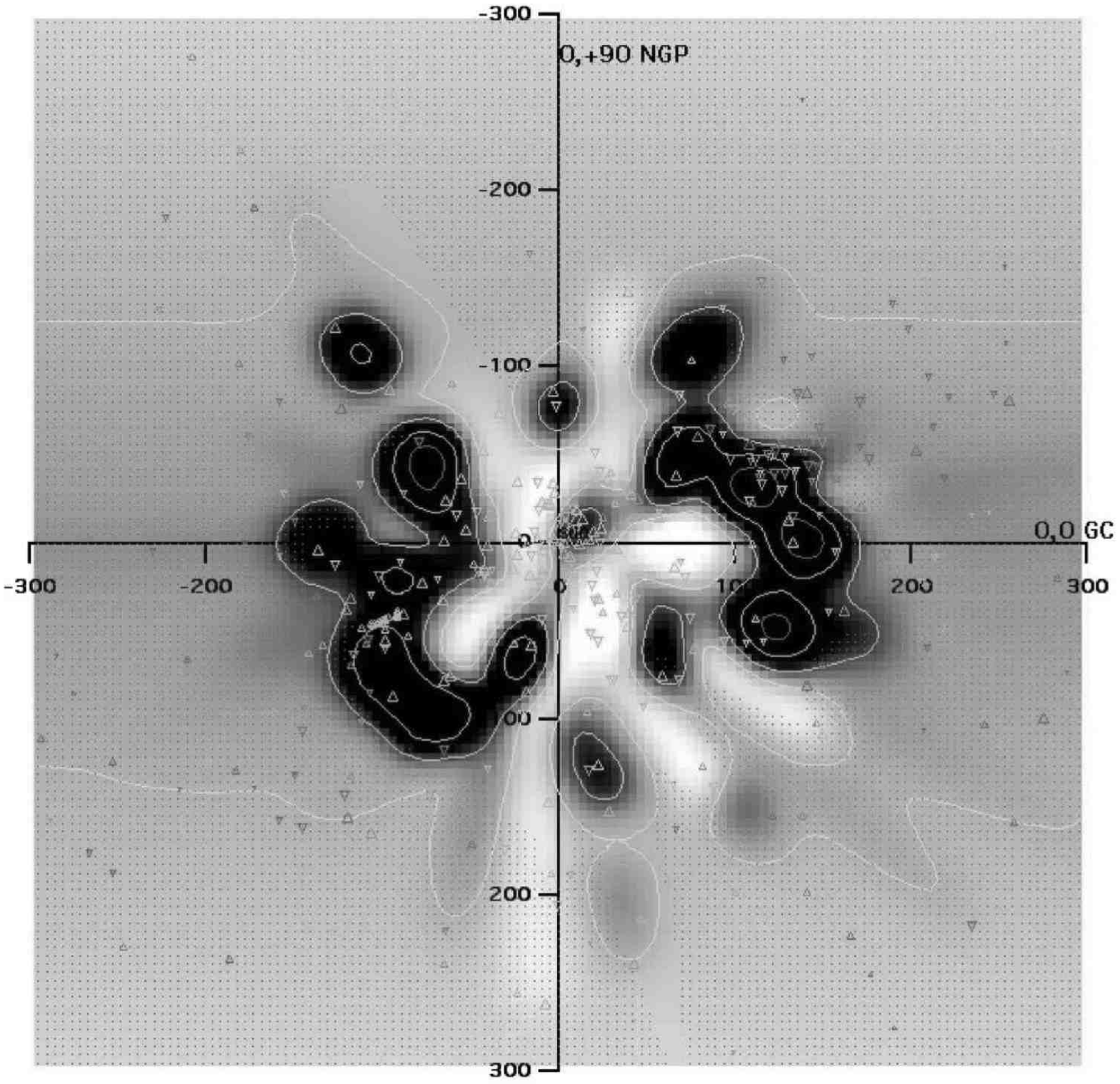}}
\caption{Plot of 3-D spatial distribution of interstellar CaII absorption within 300pc of the Sun as viewed in the meridian plane projection. Plotting symbols are the same as Figure 14.}
\label{Figure 16}
\end{figure}
\clearpage

\begin{figure}
\center
{\includegraphics[height=16cm]{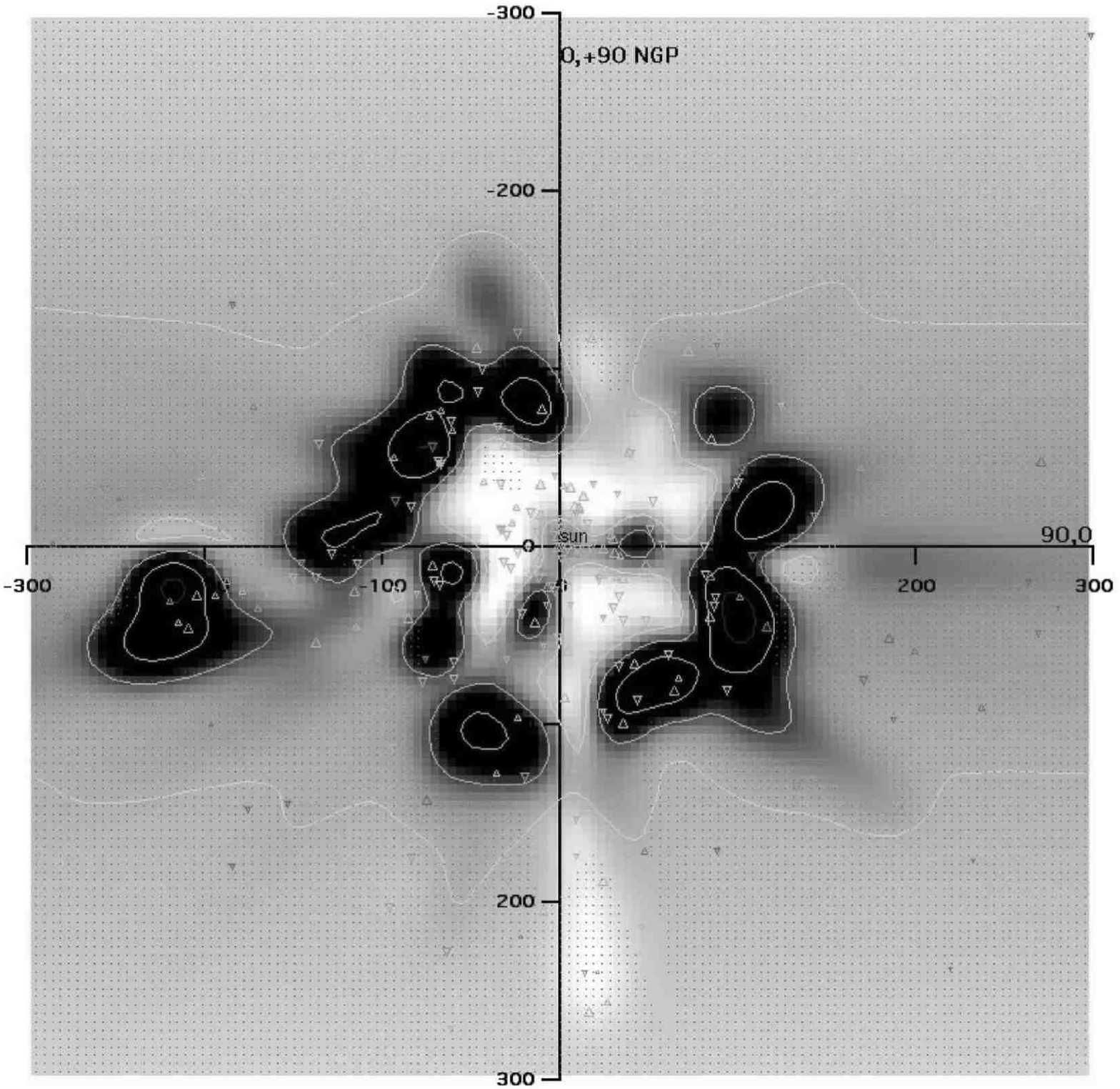}}
\caption{Plot of 3-D spatial distribution of interstellar CaII absorption within 300pc of the Sun as viewed in the rotational plane projection. Plotting symbols are the same as Figure 14.}
\label{Figure 17}
\end{figure}
\clearpage

\begin{figure}
\center
{\includegraphics[height=16cm]{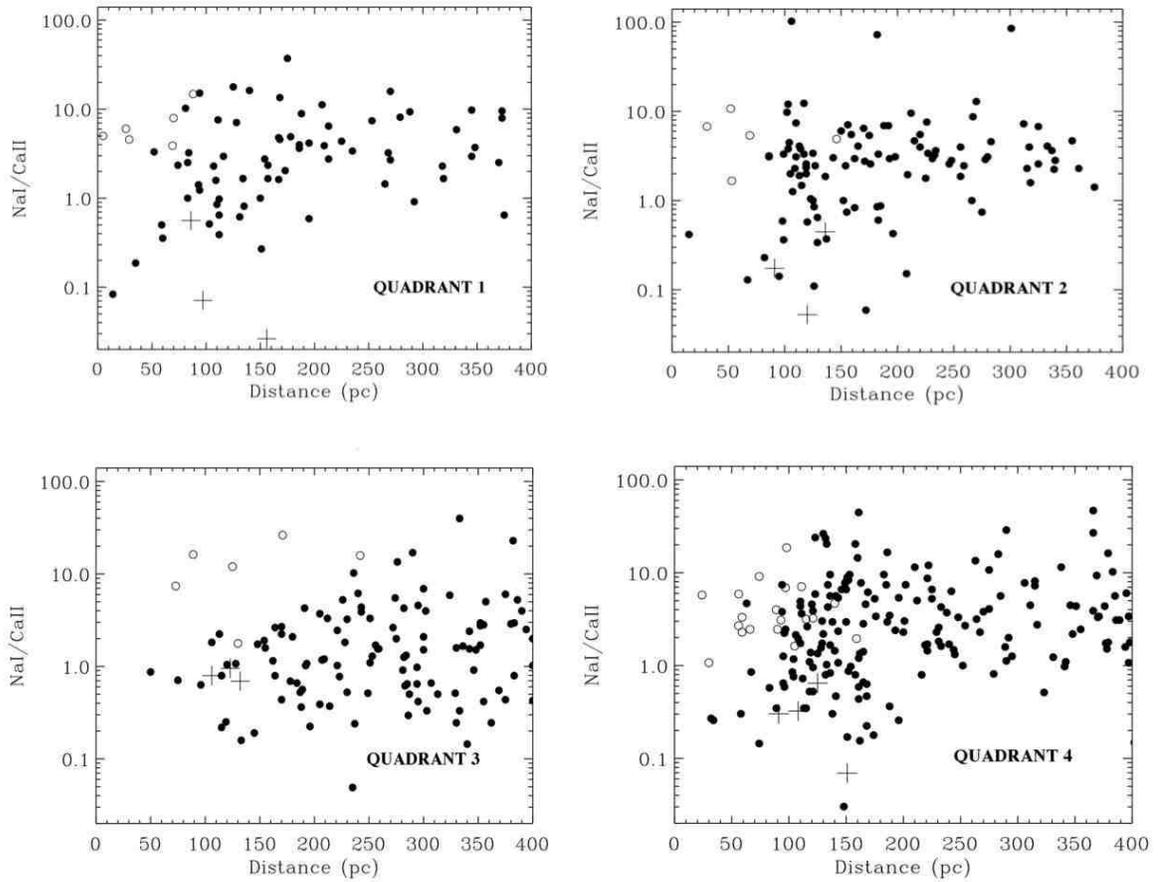}}
\caption{Plots of the column density ratio of N(NaI)/N(CaII) as a function of sight-line distance for galactic quadrants 1 - 4. Black circles are measured values, open circles are upper limits and crosses are lower limits.}
\label{Figure 18}
\end{figure}
\clearpage

\pagebreak

\begin{table*}
\begin{center}
\caption{NaI and CaII Absorption Measurements}
\begin{tabular}{lccccclcccl}
\hline
\hline
Star$_{HD}$ & l & b & distance & W$_{\lambda}$ (D2) & W$_{\lambda}$ (D1)& log N(NaI)$_{TOT}$ &Reference&  W$_{\lambda}$ (K) & log N(CaII)$_{TOT}$& Reference \\
\hline
&&&(pc)&m(\AA)&m(\AA)&cm$^{-2}$&(NaI)&m(\AA)&cm$^{-2}$&(CaII)\\
\hline
179029	&0.1	&-20	&318	&124.3	&105.4	&12.37	&ESO-2007	&53	&12.01	&ESO-2007\\
157056	&0.5	&6.6	&173	&87.0&		&11.98	& (1) 	&38.0	&11.67	&(2) \\
172016	&0.8	&-12.5	&225	&136.6	&76.20	&11.97	&ESO-2007	&25.7	&11.33	&ESO-2007\\
195599	&0.9	&-36.1	&186	&$<$2.5&		&$<$10.15&	ESO-2007	&3	&10.65	&ESO-2007\\	
203006	&1.2	&-45.0	&57		&&&&&$<$5&	$<$10.4&(3)\\
171034	&1.3	&-11.0	&685	&240.0	&175.0	&12.39&	MJO-2007&	83.4&	12.06&	MJO-2007\\
152909	&1.6	&14.4	&209	&235.0	&198.5	&12.83&	ESO-2006&	104.9&	12.24&	ESO-2006\\
154204	&1.9	&12.4	&122	&179	&142	&12.66	&(4)&&&\\			
164019	&1.9	&-2.6	&556	&&&&&	498&	12.99&	(5)\\	
161756	&2.0	&0.5		&348	&284.8	&236.6&	12.87&	ESO-2006&	99.9&	12.30&	ESO-2006\\
145570	&2.5	&28.8	&51		&6.7		&6.1&	10.67&	ESO-2008&	21.9&	11.52&	ESO-2008\\
158643	&2.5	&5.3		&131	&&	&11.05&	(6)&					20.3&	11.26&	(7)\\
165365	&2.8	&-3.7	&418	&186.4	&121.8	&12.15	&ESO-2008	&58.3	&11.91&	ESO-2008\\
180885	&2.8	&-20.7	&565	&109.8	&86.7	&12.16	&MJO-2006&&&\\			
197630	&2.9	&-38.3	&98		&12.9	&3.3		&10.71	&MJO20-06&	10.3&	11.06&	MJO-2007\\
151884	&3.2	&17.4	&268	&372.8	&348.9	&12.89	&ESO-2008&	131.8&	12.38&	ESO-2008\\
45607	&3.9	&29.7	&77		&30.0	&15.0	&11.28	&(8)		&17.6	&11.36	&LICK-2007\\
141378	&4.1	&37.3	&49	&$<$3&		&$<$10.3&	(9)&&&\\			
141569	&4.2	&36.9	&99	&&		&12.66	&(10)	&	&11.65& (10)	\\
188113	&4.2	&-27.8	&446	&199.5&152.6&12.50&ESO-2008&66.9&	12.0&	ESO-2008\\
\hline
\hline
\end{tabular}
\end{center}
\end{table*}

\end{document}